\documentclass[aip,apl,reprint]{revtex4-1}
\usepackage{times}
\usepackage{graphicx}
\usepackage{textcomp}
\usepackage{amsmath,amssymb}

\providecommand*{\ohm}{\ensuremath{\Omega}}
\providecommand*{\C}{\ensuremath{^\circ\mathrm{C}}}
\providecommand*{\micro}{\ensuremath{\text{\textmu}}}

\newcommand{\factor}{1}
\begin{document}

\title{Graphene-based electron transport layers in perovskite solar cells: a step-up for an efficient carrier collection}

\author{F. Biccari}
\affiliation{Dept. of Physics and Astronomy, University of Florence and LENS,  Via Sansone 1, I--50019 Sesto Fiorentino (FI), Italy}
\author{F. Gabelloni}
\affiliation{Dept. of Physics and Astronomy, University of Florence and LENS,  Via Sansone 1, I--50019 Sesto Fiorentino (FI), Italy}
\author{E. Burzi}
\affiliation{Dept. of Physics and Astronomy, University of Florence and LENS,  Via Sansone 1, I--50019 Sesto Fiorentino (FI), Italy}
\author{M. Gurioli}
\affiliation{Dept. of Physics and Astronomy, University of Florence and LENS,  Via Sansone 1, I--50019 Sesto Fiorentino (FI), Italy}
\author{S. Pescetelli}
\affiliation{C.H.O.S.E. (Centre for Hybrid and Organic Solar Energy), Department of Electronic Engineering, University of Rome Tor Vergata, Via del Politecnico 1, I-00133 Rome, Italy}
\author{A. Agresti}
\affiliation{C.H.O.S.E. (Centre for Hybrid and Organic Solar Energy), Department of Electronic Engineering, University of Rome Tor Vergata, Via del Politecnico 1, I-00133 Rome, Italy}
\author{A. E. Del Rio Castillo}
\affiliation{Istituto Italiano di Tecnologia, Graphene Labs, Via Morego 30, I-16163 Genova, Italy}
\author{A.~Ansaldo}
\affiliation{Istituto Italiano di Tecnologia, Graphene Labs, Via Morego 30, I-16163 Genova, Italy}
\author{E.~Kymakis}
\affiliation{Center of Materials Technology and Photonics \& Electrical Engineering Department School of Applied Technology, Technological Educational Institute (T.E.I) of Crete Heraklion, 71 004 Crete, Greece}
\author{F. Bonaccorso}
\affiliation{Istituto Italiano di Tecnologia, Graphene Labs, Via Morego 30, I-16163 Genova, Italy}
\author{A. Di Carlo}
\affiliation{C.H.O.S.E. (Centre for Hybrid and Organic Solar Energy), Department of Electronic Engineering, University of Rome Tor Vergata, Via del Politecnico 1, I-00133 Rome, Italy}
\author{A. Vinattieri}
\email[Corresponding author: ]{anna.vinattieri@unifi.it}
\affiliation{Dept. of Physics and Astronomy, University of Florence and LENS,  Via Sansone 1, I--50019 Sesto Fiorentino (FI), Italy}

\date{\today}

\begin{abstract}
The electron transport layer (ETL) plays a fundamental role in perovskite solar cells. Recently, graphene-based ETLs have been proved to be good candidate for scalable fabrication processes and to achieve higher carrier injection with respect to most commonly used ETLs. In this work we experimentally study the effects of different graphene-based ETLs in sensitized MAPI solar cells. By means of time-integrated and picosecond time-resolved photoluminescence techniques, the carrier recombination dynamics in MAPI films embedded in different ETLs is investigated. Using graphene doped mesoporous TiO$_2$ (G+mTiO$_2$) with the addition of a lithium-neutralized graphene oxide (GO-Li) interlayer as ETL, we find that the carrier collection efficiency is increased by about a factor two with respect to standard mTiO$_2$. Taking advantage of the absorption coefficient dispersion, we probe the MAPI layer morphology, along the thickness, finding that the MAPI embedded in the ETL composed by G+mTiO$_2$ plus GO-Li brings to a very good crystalline quality of the MAPI layer with a trap density about one order of magnitude lower than that found with the other ETLs. In addition, this ETL freezes MAPI at the tetragonal phase, regardless of the temperature. Graphene-based ETLs can open the way to significant improvement of perovskite solar cells.
\end{abstract}


\maketitle

\section{Introduction}

In the last decade, hybrid organic-inorganic materials with the perovskite structure, most notably CH$_3$NH$_3$Pb(I,Cl)$_3$, have become the most promising light-harvesting active layer for the implementation of high efficiency and low cost solar cells \cite{Graetzel_2014,Hodes_2013,Xiao_2016,Lee_2012,Heo_2013,You_2014,Zhou_2014,Nanni_2014}. As a matter of fact, the power conversion efficiency of perovskite solar cells (PSCs) has shown a very fast growth, increasing from 3.8\% \cite{Kojima_2009} in 2009 to 22.1\%  \cite{NREL_eff_chart} in 2016. The ease of fabrication and the high radiative efficiency have made perovskites also extremely attractive for the development of bright LEDs and lasers \cite{Veldhuis_2016,Sutherland_2016,Palma_2016,Palma_2016_2}.

Nevertheless, a major drawback originates from the poor material stability\cite{Xu_2016,Xiao_2016}: decomposition after exposure to moisture, cracks and defects generated by thermal stresses, crystal phase transition, UV-light exposure represent the principal causes of aging\cite{Xu_2016,Xiao_2016}.
Moreover, a high density of trap states, both on surfaces and in grain boundaries \cite{Xiao_2016,Draguta_2016,Wu_2015,Kim_2014}, are present in polycrystalline perovskites, used for solar cells and light emitters. 
Indeed, even though theoretical predictions show that deep trap states are not generally formed inside perovskites grains, the opposite is observed at the grain boundaries and at the surfaces\cite{Xiao_2016,Park_2016_book}.
Therefore, sophisticated passivation strategies are essential for increasing the efficiency of PSCs and light emitters \cite{Son_2016,Giordano_2016,deQuilettes_2016,Park_2016_book}.

For CH$_3$NH$_3$PbI$_3$ (MAPI) perovskites, a correlation between the solar cell efficiency and the grain morphology was recently demonstrated \cite{Shao_2016}. Results of local short-circuit photocurrent, open-circuit voltage and dark drift current within individual grains correlate these quantities to different crystal facets, as a consequence of a facet-dependent density of trap states \cite{Leblebici_2016} and it was also proven that the structural order of the electron transport layer (ETL) impacts the overall cell performance \cite{Shao_2016}. Moreover, the nature of grain boundaries was shown to affect the carrier recombination kinetics because of non-radiative pathways that would also play a role in the process of charge separation and collection \cite{deQuilettes_2016}.

It turns out that the possibility of controlling the morphology of the perovskite thin film and the understanding if different realizations of the ETL can modify the morphology of the grains are of the utmost relevance to further improvements of perovskite based solar cells.

Recently, graphene and related two-dimensional materials have been introduced in the device structure in order to improve the charge injection and/or collection at the electrodes: an enhancement of the power conversion efficiency\cite{Acik_2016} and a long-term stability\cite{Agresti_2016} was obtained.
As a matter of fact, interfaces between perovskite and transport layers have been recently demonstrated to dramatically affect the charge recombination processes and material instability within the working device.\cite{Capasso_2016}
In fact, when free charges are fast injected from perovskite to the electron transport layer, the perovskite degradation is slowed down and the non-radiative recombination is reduced\cite{Ahn_2016}. 
In particular, the insertion of graphene flakes into the mesoporous-TiO$_2$ layer (mTiO$_2$) and of lithium-neutralized graphene oxide (GO-Li) as interlayer at perovskite/mTiO$_2$ interface showed enhanced conversion efficiency and stability on both small and large area devices by demonstrating the crucial role of graphene interface engineering in perovskite-based devices\cite{Agresti_2017}. Thus, the influence of mesoscopic-graphene modified substrates onto the perovskite film need to be investigated more in details to finely control the photovoltaic performance of complete devices.

Given the typical size of the grains (a few hundreds of nanometers), high resolution techniques as TEM (Transmission Electron Microscopy), AFM (Atomic-Force Microscopy), SNOM (Scanning Near-field Optical Microscopy), etc.~are employed to investigate the grain morphology but a difficult task is to correlate physical properties at the nanoscale with the device performance measuring, for instance, the $I$-$V$ curve of the cell\cite{Leblebici_2016}.
Thus, it would be of  extreme relevance to extract information on the active film morphology with much easier techniques on a length scale of tens/hundreds of microns and even larger so to assess the homogeneity of the film deposition and the reliability of the synthesis protocol and post-deposition treatments. Moreover, the high spatial resolution analysis can explore a limited thickness of the film, and therefore it appears difficult to get information in the case of a real device.
Photoluminescence (PL) spectroscopy is an effective tool to investigate the film quality: in fact, from the comparison of samples with different ETLs, it is possible to extract quantitative information on the carrier capture and transport and, by the spectral shape, identify the crystalline phase of the active layer and evaluate the density of traps/defects.
Moreover, given the limited thickness (few hundreds of nm) of the perovskite film in solar cells and given the steep behavior of the absorption coefficient in MAPI\cite{Loper_2015}, by varying the excitation wavelength in the range 300-700\,nm, it is possible to probe thicknesses of the film from a hundred of nm to the whole film layer.

In this paper we aim to establish the effects of different graphene-based
ETLs in sensitized MAPI solar cells. In particular we will study the ETL
effects on the carrier collection efficiency and on the MAPI morphology along
the thickness.
By picosecond time-resolved measurements we correlate the carrier recombination dynamics to the crystalline quality of the active material in presence/absence of the ETL. We will find an increase of the electron collection efficiency up to a factor 3 with respect to standard mTiO$_2$.
Taking advantage of the absorption coefficient dispersion, we are able to assess the film morphology along the thickness. In fact, by tuning the excitation wavelength, we investigate a thickness range from 150 to 400\,nm and we can highlight the morphology changes induced by different ETLs.
Our results will indicate that, when a graphene doped mesoporous TiO$_2$ (G+mTiO$_2$) with the addition of a GO-Li interlayer is used as ETL, the morphology of the MAPI film embedded in the mesoporous layer is frozen in the tetragonal phase, regardless of the temperature. In addition, the defect concentration is about one order of magnitude lower than that found with the other ETLs.

\section{Results and discussion}

\begin{figure}
\includegraphics[width=\factor\columnwidth]{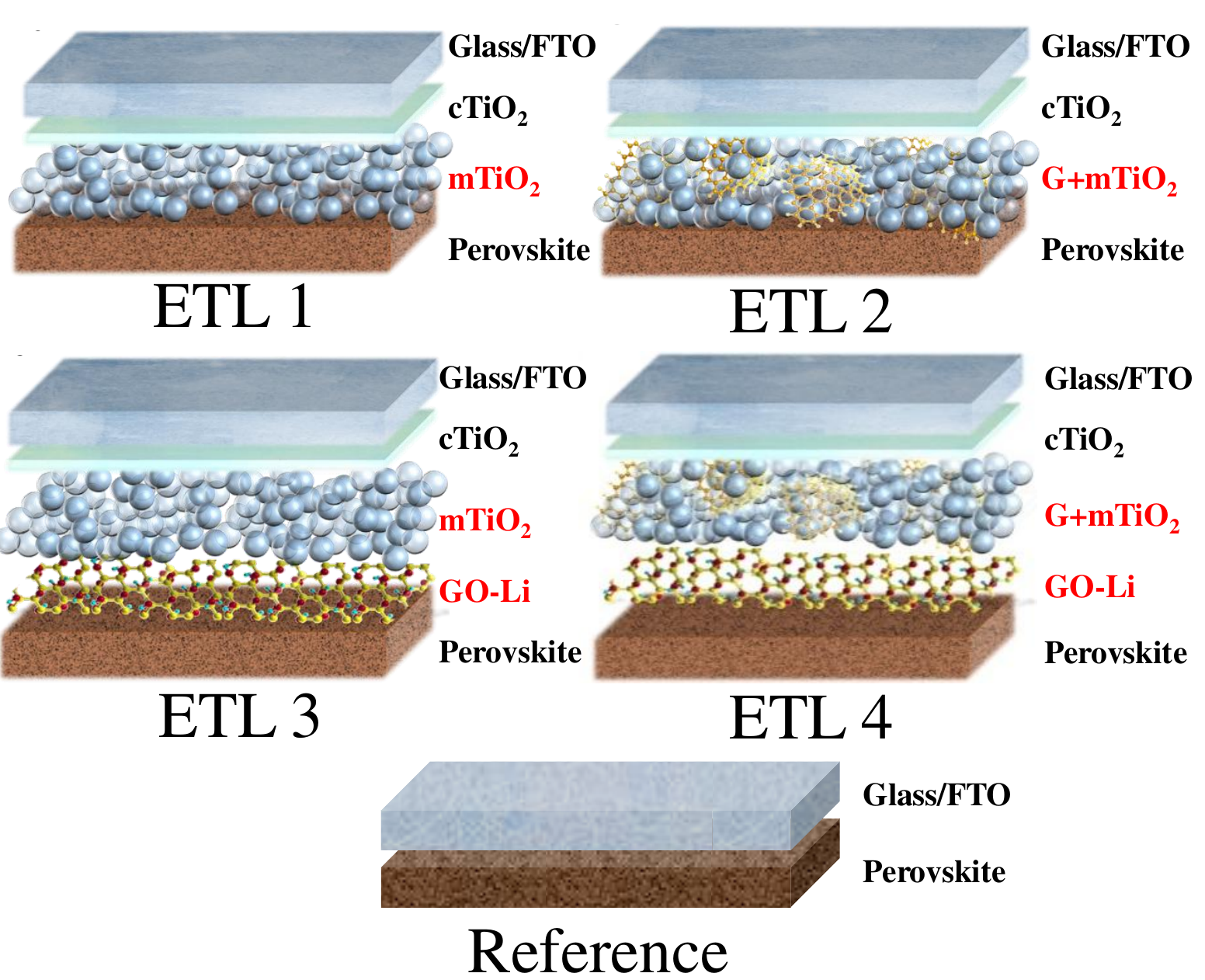}
\caption{Structures of the investigated samples. The ETLs are indicated in red.}
\label{Fig1}
\end{figure}

Four types of samples were prepared using different combinations of ETLs: mTiO$_2$, G+mTiO$_2$, mTiO$_2$ plus GO-Li interlayer and G+mTiO$_2$ plus GO-Li interlayer. 
The different sample structures are schematically shown in Fig.\,\ref{Fig1} and listed in Tab.\,\ref{Tab1}. For clear comparison a sample of MAPI on FTO without ETL (reference sample) was also investigated. 
Details on the sample preparation are reported in the Experimental Section.
It is worth noting that PL analysis was carried out on simple photo-electrodes, lacking of the hole collecting layer and the bottom contact: this allows us to focus only on the electron collection and transport after the electron-hole pairs creation due to the photon absorption. Moreover, PL experiments can be realized illuminating the samples either from the perovskite film side (side A) or the FTO side (side B). Varying the excitation wavelength and the excitation side, we can differently penetrate into the MAPI film and selectively probe spatial regions few tens of nm near the ETL or far from it.

\begin{table}[b]
\caption{List of the investigated samples and description of their corresponding ETLs.}
\label{Tab1}
\addtolength{\tabcolsep}{3.8mm}
\begin{tabular*}{\columnwidth}{l@{\hspace{22mm}}l}
\hline
Sample    & ETL  \\
\hline
Reference & No ETL       \\
ETL 1         & mTiO$_2$   \\
ETL 2         & G+mTiO$_2$  \\
ETL 3         & mTiO$_2$ plus GO-Li\\
ETL 4         & G+mTiO$_2$ plus GO-Li \\
\hline
\end{tabular*}
\end{table}

\begin{figure}
\includegraphics[width=\factor\columnwidth]{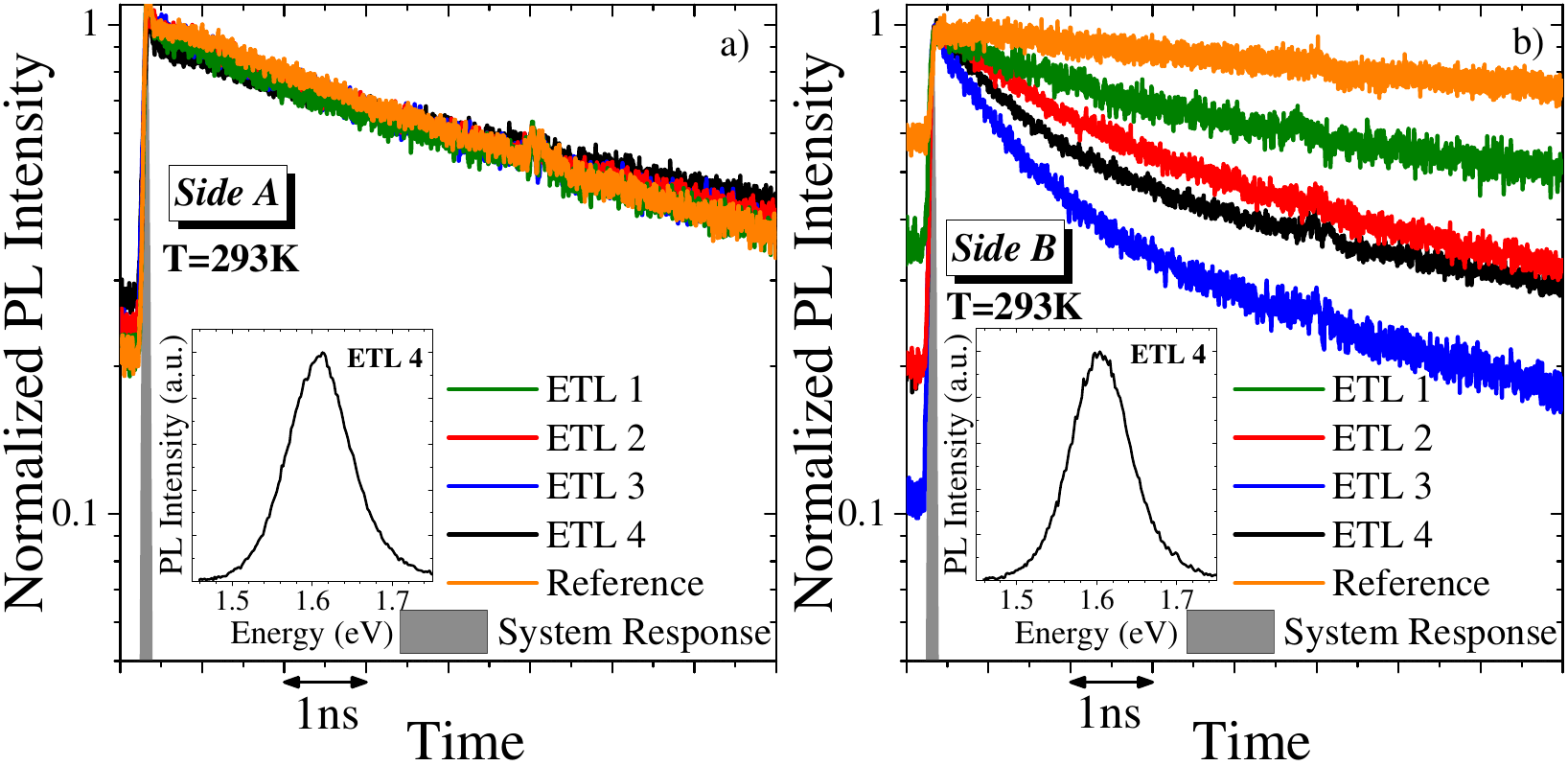}
\caption{PL decay (at the PL peak energy) and PL spectra (in the inset) at room temperature, after ps excitation at 2.06\,eV with an average intensity of 10\,W/cm$^2$. a) From side A. b) From side B.}
\label{Fig2}
\end{figure}

In Fig.\,\ref{Fig2} PL decays at room temperature, after excitation with photons of 2.06\,eV, are compared for the different samples, while in the inset a PL spectrum at room temperature is shown: typical spectra of the tetragonal phase are observed\cite{Milot_2015,Dar_2016}. 
By exciting from side A (Fig.\,\ref{Fig2}a), the PL decays are identical for all samples and no effect related to the presence of the ETL is detected. 
In contrast there is a significant difference in the PL decay by exciting from side B depending on the sample, as shown in Fig.\,\ref{Fig2}b: faster decays are observed in presence of ETL, especially when graphene and/or GO-Li are used.
It is worth noting that the PL decay of reference sample excited from side B is slower with respect to the decays of side A. This can be attributed to non-radiative states at the surface of the uncovered perovskite film (side A).

To extract information about the efficiency in the carrier injection from perovskite to ETL, we first fitted the decays of Fig.\,\ref{Fig2}b with a double exponential\cite{Son_2016,Bi_2016}, taking into account the laser pulse repetition period \cite{Warren_2013}. The fitting function $I(t)$ can be written as
\begin{equation}
I(t) = f(t)+g(t)
\end{equation}
where $f(t)$ is the original double exponential and $g(t)$ the correction term due to the intrinsic periodic nature of TCSCP measurements\cite{Warren_2013}:
\begin{equation}
f(t)=\Theta(t-t_0)[Ce^{-(t-t_0)/\tau_1}+ 
(1-C)e^{-(t-t_0)/\tau_2}]
\end{equation}
\begin{equation}
g(t)=C\frac{e^{-(t-t_0)/\tau_1}}{e^{T/\tau_1}-1}+ 
(1-C)\frac{e^{-(t-t_0)/\tau_2}}{e^{T/\tau_2}-1} 
\end{equation}
In the previous equations $\theta$(t) is the Heaviside function, $\tau_1$ and $\tau_2$ are the decay time constants, $T$ is the laser pulse repetition period (13.15\,ns), $C$ is the contribution of $\tau_1$ exponential to the fit and $t_0$ is a constant. 
The results obtained by the fitting procedure are reported in Tab.\,\ref{Tab2}.
Inserting graphene-based ETLs in the samples reduces $\tau_1$ from 25\,ns to 15\,ns while the reduction of $\tau_2$ depends on the ETL.
In the literature\cite{Son_2016,Bi_2016} the longer decay constant ($\tau_1$) is ascribed to the radiative recombination in MAPI while the shorter decay constant can be attributed to the carrier removal from MAPI layer towards the ETL.

To get rid of the local inhomogeneities in the samples, which can give rise to variation of the TI-PL intensity and considering that the PL time evolution does not depend on the detection spot, we estimate the TI-PL intensity from the PL decay. It turns out that we can express the integrated PL intensity ($I_\mathrm{PL}$) as
\begin{equation}
I_\mathrm{PL}=\eta P,
\end{equation}
where $\eta$ is the radiative efficiency and $P$ is the pump intensity, equals to 10\,W/cm$^2$ for all the TR-PL measurements.
Assuming a unitary radiative efficiency for the reference sample, we can evaluate the change in $\eta$ for the different ETLs (see Tab.\,\ref{Tab2}). Lower $\eta$ is obtained in case of ETLs 2, 3 and 4: thus the insertion of graphene and/or GO-Li interlayer improves the electron capture from perovskite to ETL by a factor between two and three with respect to ETL 1.
We want to remark that this result does not necessarily imply an increase of the solar cell short circuit current density ($J_\mathrm{sc}$) of the same amount. In fact hole collection by the hole transport layer\cite{Pydzinska_2016} and non-radiative recombination in the ETL have to be taken into account. However, as shown in the Supporting Information, a significant increase of $J_\mathrm{sc}$ is measured for complete cell when ETL 4 is used.

\begin{table}
\caption{Results of the fits of data of Fig.\,\ref{Fig2}b: $\tau_1$ and $\tau_2$ are the decay time constants, $C$ is the contribution of $\tau_1$ to the fit and $\eta$ is the radiative efficiency.}
\label{Tab2}
\addtolength{\tabcolsep}{3.8mm}
\begin{tabular*}{\columnwidth}{lcccc}
\hline
Sample    & $\tau_1$ (ns) & $\tau_2$ (ns) & $C$  &$\eta$ \\
\hline
Reference & 25       & --    & 1.00 &1 \\
ETL 1         & 25       & 2.05   & 0.43 & 0.48 \\
ETL 2         & 15       & 1.99   & 0.36 & 0.27\\
ETL 3         & 15       & 1.24   & 0.20 &0.16\\
ETL 4        & 15       & 1.30   & 0.34 &0.24\\
\hline
\end{tabular*}
\end{table}

\begin{figure}
\includegraphics[width=\factor\columnwidth]{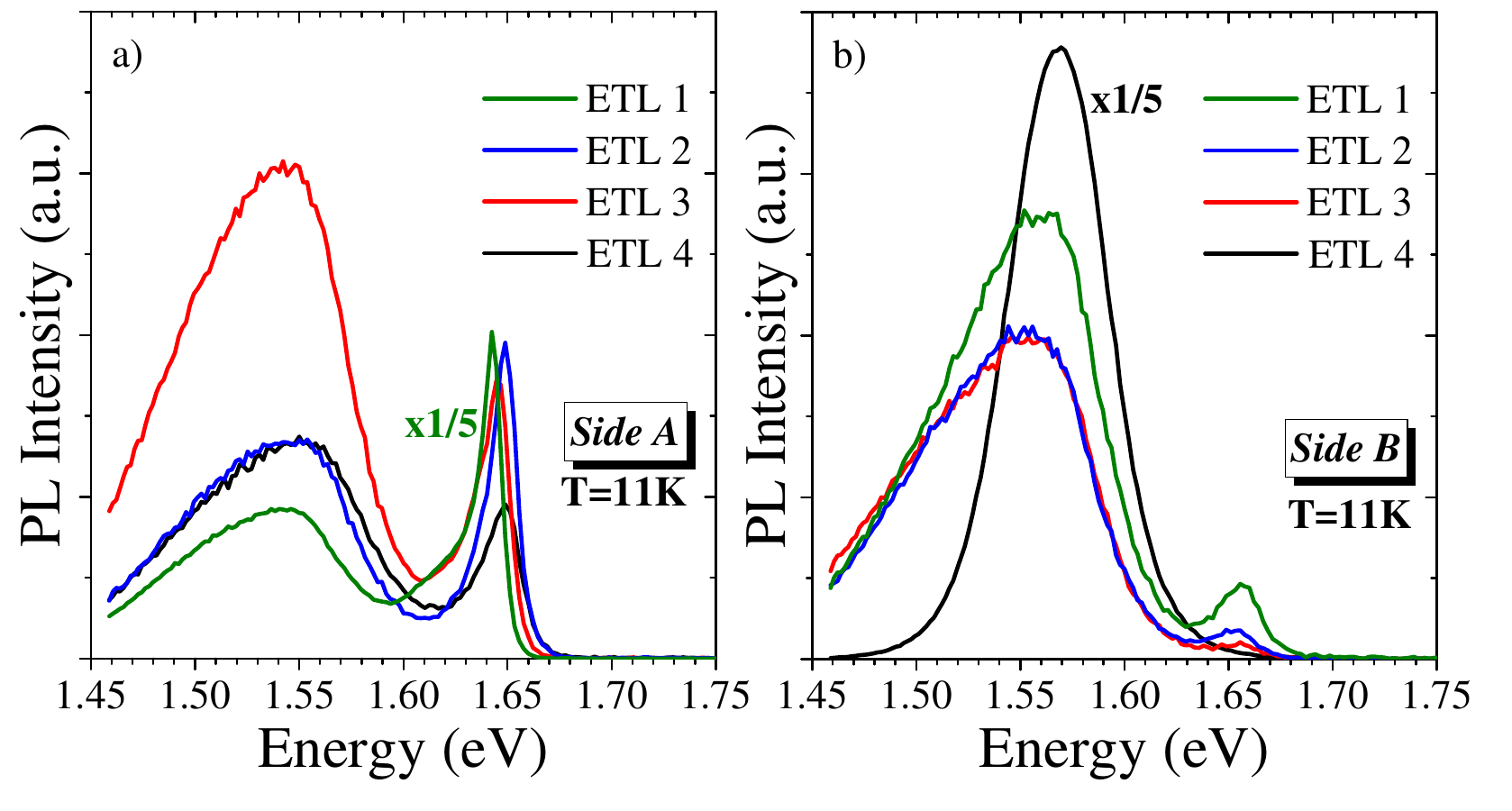}
\caption{TI-PL spectra at 11\,K for the various samples after excitation at 2.06\,eV  with an average intensity of 10\,W/cm$^2$. a) From side A. b) From side B.}
\label{Fig3}
\end{figure}

More insight in the role of the ETL can be obtained by PL spectra at low temperature ($T=11$\,K). In Fig.\,\ref{Fig3} PL spectra, obtained by exciting the samples from side A (Fig.\,\ref{Fig3}a) and side B (Fig.\,\ref{Fig3}b), are reported.
The PL spectra from side A show two peaks. As expected for $T < 150$\,K in MAPI perovskite \cite{Milot_2015,Kong_2015}, the peak at about 1.65\,eV is attributed to the orthorhombic phase of MAPI. But the major contribution to the spectra comes from the other peak, centered at 1.55\,eV. This emission is likely due to the sum of two contributions: the radiative recombination arising from the residual tetragonal phase at 1.56\,eV and the radiative recombination from localized states below 1.52\,eV. In the literature, these localized states are identified with radiative traps\cite{Wu_2015,Fang_2015} or, recently, to methylammonium-disordered domains in orthorhombic phase of MAPI\cite{Dar_2016}.
We want to remark that, in both interpretations of the low energy side emission as radiative traps or disordered domains, a carrier localization is present. More relevant is the fact that the low energy states are radiative and do not produce a loss of photogenerated carriers.

Considering that the absorption length of the MAPI film at about 2\,eV is roughly 200\,nm \cite{Loper_2015} and that the thickness of the perovskite layer in our samples is about 350\,nm (see Fig.\,S5 in the Supporting Information), we can conclude that the emission, exciting from side A, comes mostly from the MAPI film and that no effect related to the presence of the ETL is detected. 
On the contrary, the  excitation from side B can reveal the ETL effect on the MAPI. As a matter of fact, relevant differences are observed between the emissions from side B (Fig.\,\ref{Fig3}b). First of all we observe a smaller  contribution of the orthorhombic phase for all the samples with respect to the excitation from side A. Moreover, in the case of GO-Li plus G+mTiO$_2$ (ETL 4), we detect a strong reduction of the radiative traps at 1.52\,eV and the dominance of the emission from the tetragonal phase at 1.56\,eV.

Such results suggest that an incomplete phase transition occurs for the perovskite wrapped into the mesoporous ETL layer. In particular, in the case of ETL 4, where the PL lineshape shows a negligible low energy tail and a smaller linewidth with respect to the other samples, we can argue that the crystallization of the MAPI film is very good, as also confirmed by the scanning electron microscopy (SEM) image reported in Fig.\,S4 of the Supporting Information. However, the interaction of MAPI and ETL inhibits the phase transition, at least for a thickness of about 200\,nm (see SEM cross section in Fig.\,S5 of the Supporting Information). In order to confirm and test our hypothesis we performed PL measurements as a function of temperature, excitation wavelength and excitation power.

\begin{figure}
\includegraphics[width=\factor\columnwidth]{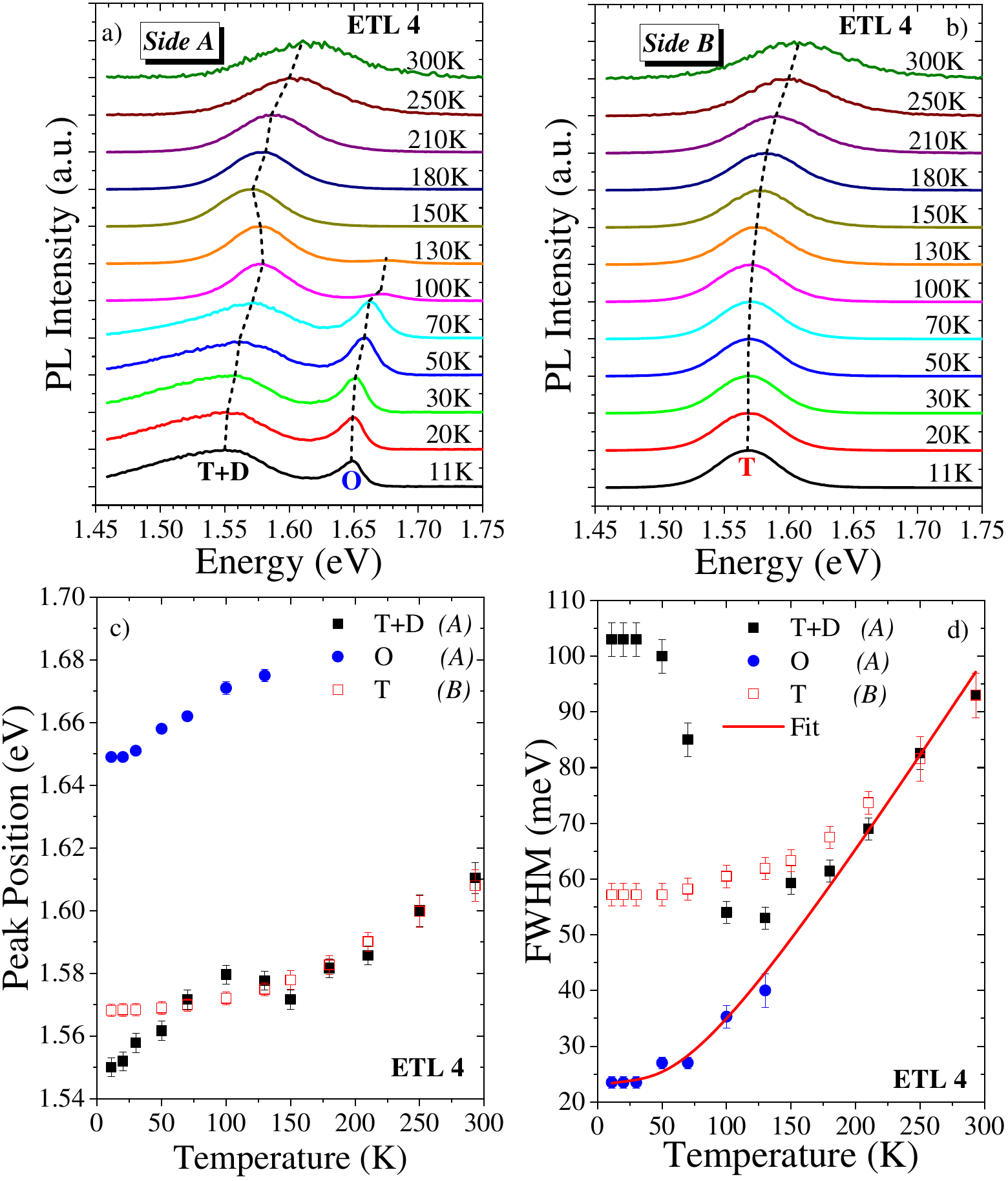}
\caption{Temperature-dependent measurement on ETL 4 by exciting with photon energy of 2.06\,eV with an average intensity of 10\,W/cm$^2$. The labels D, T and O stand for defects (trap states), tetragonal and orthorhombic, respectively. a) PL spectra from 10 to 300\,K after excitation on side A. b) PL spectra from 10 to 300\,K after excitation on side B. c) Position of the PL peaks of the spectra in Fig.\,\ref{Fig4}a and Fig.\,\ref{Fig4}b as a function of temperature. The excitation side is indicated in brackets. d) FWHM of the PL peaks of Fig.\,\ref{Fig4}a and Fig.\,\ref{Fig4}b as a function of temperature. The excitation side is indicated in brackets. Red solid line shows the fitting of FWHM through Eq.\,\eqref{Eq_Gamma}.}
\label{Fig4}
\end{figure}

In Fig.\,\ref{Fig4}a we report the emission spectra of ETL 4 exciting from side A, varying the sample temperature from 10 to 300\,K. As already shown before, at low temperature we observe two bands, one at 1.65\,eV corresponding to the orthorhombic phase and one at about 1.55 eV corresponding to the sum of the emission from optically active trap states and the emission from a residual tetragonal phase.
As expected\cite{Kong_2015,Fang_2015,Dar_2016,Wright_2016}, increasing the temperature, the orthorhombic phase emission shifts at higher energy (see Fig.\,\ref{Fig4}c), showing a monotonic increase of its full width at half maximum (FWHM) (see Fig.\,\ref{Fig4}d), and it disappears above 150\,K, where the phase transition of MAPI from orthorhombic to tetragonal phase occurs.
By increasing the temperature, the low energy band shows instead the typical S shape both in the peak emission energy (see Fig.\,\ref{Fig4}c) and its FWHM (see Fig.\,\ref{Fig4}d). This is an indication of the phase transition from orthorhombic to tetragonal phase, with the concurrent lower contribution of the traps.
Above 150\,K the PL spectrum has only one peak, arising from the tetragonal phase emission, which continues to monotonically blue shift increasing the temperature (see Fig.\,\ref{Fig4}c), as expected\cite{Kong_2015,Fang_2015,Dar_2016,Wright_2016}. 
Apart from different relative weights of the emission bands, for all samples we find PL spectra very similar to the one of the sample with ETL 4 when the excitation is performed from side A.

A very similar trend with temperature, as shown in Fig.\,\ref{Fig4}a for ETL 4, is found for ETL 1, 2 and 3, irrespective of the excitation side, and this trend is commonly reported in the literature\cite{Dar_2016}. On the contrary, relevant differences are observed in case of ETL 4 exciting from side B (Fig.\,\ref{Fig4}b). By increasing the temperature, the PL spectrum shows a single band, corresponding to the tetragonal phase, with a monotonic increase of the emission energy (see Fig.\,\ref{Fig4}c) and the FWHM (see Fig.\,\ref{Fig4}d). Such behavior indicates  that the  MAPI film embedded in the mesoporous ETL side remains in the tetragonal phase even down to 10\,K.

The FWHM of the PL bands as a function of temperature can be fitted taking into account the temperature-independent inhomogeneous broadening and the interaction between carriers and acoustic and longitudinal optical (LO) phonons\cite{Dar_2016,Wright_2016}, using the following equation:
\begin{equation}
\Gamma(T)=\Gamma_0+\gamma_{\mathrm{ac}}T + \frac{\gamma_{\mathrm{LO}}}{e^{E_{\mathrm{LO}}/k_\mathrm{B}T}-1},
\label{Eq_Gamma}
\end{equation}
where $\Gamma_0$ is the inhomogeneous broadening, $\gamma_{\mathrm{ac}}$ and $\gamma_\mathrm{LO}$ are the acoustic and LO phonon-carrier coupling strengths, respectively, and $E_\mathrm{LO}$ is the LO phonon energy.
We fitted the FWHM data extracted from the PL from side A: in particular we considered the orthorhombic phase from 10\,K to 150\,K and the tetragonal phase from 150\,K up to room temperature. If the two set of data can be fitted with a single function, we can conclude that the acoustic and optical phonons causing the PL line broadening have very similar energies for the two phases. The solid line in Fig.\,\ref{Fig4}d shows the best fitting curve and a good agreement is reached between the data and the model with the fitting parameters $\Gamma_0 = (23\pm1)$\,meV, $\gamma_\mathrm{ac} = (30\pm5)$\,\micro eV/K, $\gamma_\mathrm{LO} = (75\pm5)$\,meV, $E_\mathrm{LO} = (19\pm1)$\,meV. These values  well agree with  data in the literature \cite{Dar_2016,Wright_2016}.

\begin{figure}
\includegraphics[width=\factor\columnwidth]{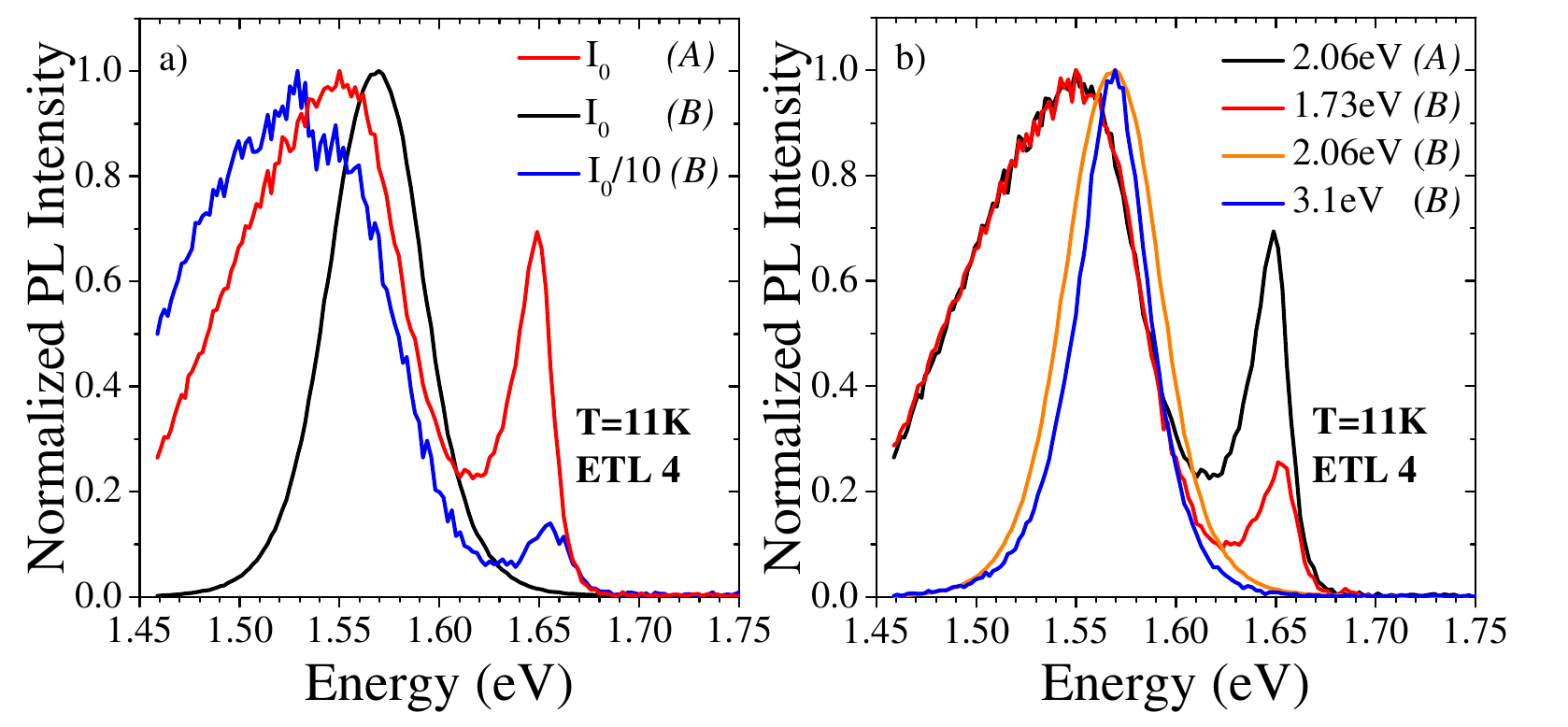}
\caption{a) Normalized PL spectra of ETL 4 at $T=11$\,K, after excitation at 2.06\,eV, for different excitation densities and excitation sides (in brackets). $I_0$ corresponds to an average intensity of 10\,W/cm$^2$. b) Normalized PL spectra of ETL 4 at $T=11$\,K for different excitation photon energies and excitation sides (in brackets).}
\label{Fig5}
\end{figure}

A further evidence proving that the emission band at low temperature in ETL 4 must be attributed to a tetragonal crystalline phase of MAPI with good quality, is given in Fig.\,\ref{Fig5}a. We report two spectra, already shown before, from side A (red curve) and side B (black curve), acquired at the same average excitation intensity $I_0=10$\,W/cm$^2$.
Decreasing the excitation intensity by one order of magnitude, the emission from side B changes completely, showing a spectrum (blue curve) similar to the side A (red curve), with the dominant contribution of the traps and a small signal from the orthorhombic phase. This is explained by the fact that, lowering the power density, only the trap energy levels close to the band gap tail are filled.
In addition, the PL spectrum of the sample ETL 4, obtained by exciting from side B with an intensity of $I_0/10$ (blue curve in Fig.\,\ref{Fig5}a), is very similar to the spectra of the ETLs 1, 2 and 3 (Fig.\,\ref{Fig3}b) by exciting from side B with an higher intensity $I_0$. This means that the trap density in the mesoporous region of MAPI in ETL 4 is lower, of about one order of magnitude, with respect to the other samples. This result agrees with the better crystal quality found by SEM analysis for ETL 4 (see Fig.\,S4 in the Supporting Information).

As stated previously, to confirm that the crystalline nature of the film changes when in contact with the ETL, in particular in presence of GO-Li plus G+mTiO$_2$, we probed the ETL 4 along the thickness exploiting the different absorption coefficient of MAPI varying the excitation photon energies.
We spanned the range from 1.73 to 3.1\,eV, where FTO and cTiO$_2$ have a low and nearly constant absorption.
The result of this experiment, performed at 11\,K, is reported in Fig.\,\ref{Fig5}b. Let us focus on side B. At high photon energy excitation, 3.1\,eV (blue curve) and 2.06\,eV (orange curve), the MAPI is excited for a few tens of nanometers close to the ETL and, as already observed above, the spectra show only the tetragonal phase (which should not be observed at this temperature since the only stable phase below 150\,K is the orthorhombic). Decreasing the excitation photon energy down to 1.73\,eV, the absorption coefficient of MAPI decreases and therefore the sample is excited more uniformly in depth. The resulting spectrum (red curve) shows an increase of the contribution of the orthorhombic phase.
For comparison, a spectrum from side A with an excitation of 2.06\,eV is reported (black curve). The comparison of this spectrum with that at 1.73\,eV from side B (red curve) shows, apart from a difference in the intensity of the orthorhombic phase, exactly the same contribution from the radiative traps and the tetragonal phase. The observed behavior proves that the crystalline nature of MAPI at low temperature is influenced by the interaction with the ETL, which inhibits the MAPI phase change into the orthorhombic form.

Our results demonstrate a substantial improvement of the active layer morphology of ETL 4 with an efficient carrier capture from the ETL.
This is confirmed by a remarkable increase in power conversion efficiency (PCE) for complete devices, obtained by $I$-$V$ characterization (see Fig.\,S6, S7 and Tab.\,S1 in the Supporting Information), which is mainly ascribed to an improved short circuit current density ($J_\mathrm{SC}$).
%Recently\cite{Agresti_2017}, $I$-$V$ results on similar samples indicate an increase in efficiency and stability of the cell when graphene flakes are used to dope the mTiO$_2$ layer with a GO-Li interlayer, as ETL, even in large area device, exceeding 50\,cm$^2$.
%This increase of the efficiency can now be explained as the result of a better structural quality of the MAPI film embedded in the mTiO$_2$ layer when GO-Li is added and graphene is used as dopant.

\section{Conclusions}

We investigated the effects of different graphene-based ETLs in sensitized MAPI photoelectrodes. In particular we have compared four different samples with the following ETLs: mTiO$_2$, G+mTiO$_2$, mTiO$_2$ plus GO-Li interlayer and G+mTiO$_2$ plus GO-Li interlayer. 
We have studied the ETL effects on the carrier collection efficiency and on the MAPI morphology and quality along the thickness.
In presence of ETL, we found faster PL decays by exciting on FTO side with respect to the MAPI side which is explained by efficient electron removal from MAPI layer due to the ETL. In particular an increase of the electron collection efficiency up to a factor 3 with respect to standard mTiO$_2$ is reported.

Moreover, the MAPI layer embedded in G+mTiO$_2$ plus GO-Li ETL shows a crystalline quality much better than the other samples, with a trap density about one order of magnitude lower.
Exploiting the dispersion of the MAPI absorption coefficient, we could probe the sample along the thickness, finding that the morphology of the MAPI film embedded in the G+mTiO$_2$ plus GO-Li ETL is frozen in the tetragonal phase, regardless of the temperature.
Moreover, the observed morphology improvement of the MAPI encapsulated in the mTiO$_2$ plus GO-Li layer supports the increased efficiency measured for the complete devices.

Finally, our results show that graphene based ETLs significantly improve both the carrier collection and the crystalline quality of the active material, opening new routes to the development of efficient and stable MAPI solar cells.

\section{Experimental section}

\textbf{Sample preparation.} 
Solar cells photoelectrodes were prepared on Fluorine-doped Tin Oxide (FTO) conductive glass (Pilkington TEC 8, 8\,\ohm/$\square$, 25\,mm${}\times{}$25\,mm). 
The substrates were cleaned in an ultrasonic bath, using three sequential steps: detergent with de-ionized water, acetone and 2-Propanol (10\,min for each step).
The substrates were covered by a compact layer of TiO$_2$ (cTiO$_2$).
A solution of acetylacetone (2\,mL), titanium diisopropoxide (3\,mL) and ethanol (45\,mL) was deposited onto the FTO substrates by Spray Pyrolysis Deposition at 450\,\C. The final thickness of the cTiO$_2$ layer was measured about 50\,nm by a Dektak Veeco 150 profilometer.

The mTiO$_2$ layer was obtained starting by an ethanol solution of 18NR-T titania paste (Dyesol) dissolved in pure ethanol (1:5 by weight), stirred overnight. 
The graphene-doped mTiO$_2$ was obtained by adding sonicated graphene ink (1\% in vol.),
prepared by dispersing 5\,g of graphite flakes (+100 mesh, $\ge$75\%, Sigma Aldrich) in 500\,mL of N-methyl-2-pyrrolidone, NMP (Sigma Aldrich). The initial dispersion was ultrasonicated (VWR) for 6 hours and subsequently ultracentrifuged using a SW32Ti rotor in a Beckman-Coulter Optima XPN ultracentrifuge at 10000\,rpm ($\sim$12200 g) for 30 mins at 15\,$^\circ$C. After ultracentrifugation, the upper 80\% supernatant was extracted by pipetting. The concentration of the graphitic flakes is calculated from the OAS (see Fig.\,S1 in the Supporting Information), giving a concentration of 0.25\,g L$^{-1}$. The morphology of the flakes, i.e., lateral size and thicknesses are characterized by transmission electron microscopy (TEM) and atomic force microscopy (AFM), respectively (see Fig.\,S2 in the Supporting Information) giving a lateral size distribution of 150\,nm and thickness of 1.7\,nm. Raman spectroscopy data are found in the Supporting Information (see Fig.\,S3).
Both standard mTiO$_2$ and G+mTiO$_2$ dispersions were sonicated 10\,min prior to be spin-coated in air at 1700\,rpm for 20\,s onto the cTiO$_2$ surface, followed by a calcination step at 450\,\C\ for 30\,min.

The GO-Li interlayer was realized by spin coating (2000\,rpm for 10\,s) 200\,\micro L of GO-Li dispersion in ethanol/H$_2$O (3:1) prepared as reported in Ref.\,\onlinecite{Agresti_2016b}. After the deposition, the substrates were annealed at 110\,\C\ for 10\,min.

The photo-electrodes were completed by depositing the perovskite active layer in dry conditions (relative humidity less than 30\%) by a double step method: a lead iodide solution (PbI$_2$ in N,N-dimethylformamide, 1\,M, heated at 70\,\C) was spin coated at 6000\,rpm for 10\,s on heated substrates (50\,\C) which were then dipped into a CH$_3$NH$_3$I (Dyesol) in anhydrous 2-propanol solution (10\,mg/mL) for 15\,min. Finally, the samples were heated at 80\,\C\ for 20\,min in air. The MAPI absorbing layer has a typical thickness of 350\,nm with a perovskite penetration into mesoporous layer of about 200\,nm.
The samples were not encapsulated. 
Complete solar cells were realized with the same structure of the investigated samples to compare the PL results with the current-voltage ($I$-$V$) characteristics.
In the case of complete devices doped spiro-OMeTAD (73.5\,mg/mL) in chlorobenzene solution doped with tert-butylpyridine (TBP 26.77\,\micro L/mL), lithium bis(trifluoromethanesulfonyl)imide (LiTFSI 16.6\,\micro L/mL), and cobalt(III) complex (FK209 from Lumtec, 7.2\,\micro L/mL) is spin coated (2000\,rpm for 20\,s) onto the tested photo-electrodes. The final devices are completed by Au counter-electrode thermal evaporation (100\,nm). $I$-$V$ characteristics are recorded under AM1.5G solar simulator Solar Constant from KHS at 1000\,W/m$^2$ (1\,sun). Results on the solar cells are shown in the Supporting Information.

\textbf{Optical Absorption Spectroscopy (OAS).}
The OAS of the as-produced inks was carried out in the 300-1000\,nm range with a Cary Varian 5000i UV-vis-NIR spectrometer. The absorption spectra was acquired using a 1\,mL quartz glass cuvette. The ink is diluted to 1:7 in NMP. The NMP solvent baseline was subtracted. The concentration of graphitic flakes is determined from the extinction coefficient at 660\,nm, using $A = \alpha l c$ where $l$ [m] is the light path length, $c$ [gL$^{-1}$] is the concentration of dispersed graphitic material, and $\alpha$ [Lg$^{-1}$m$^{-1}$] is the absorption coefficient, with $\alpha \sim 1390$\,Lg$^{-1}$m$^{-1}$ at 660\,nm.\cite{Lotya_2009}

\textbf{Transmission electron microscopy (TEM).}
The exfoliated flakes morphology is characterized by using a TEM JOEL JEM 1011, using an acceleration voltage of 100\,kV. The sample preparation was performed diluting the ink  in NMP (1:10). 20\,\micro L of the diluted sample were drop cast on copper grids (200 mesh), and dried in vacuum overnight.  Statistical analyses are fitted with log-normal distributions.

\textbf{Atomic force microscopy (AFM).}
The dispersions are diluted 1:30 in NMP. 100\,\micro L of the dilutions are drop-casted onto Si/SiO$_2$ wafers. AFM images are acquired with Bruker Innova AFM in tapping mode using silicon probes (frequency = 300\,kHz, spring constant = 40\,Nm$^{-1}$). Statistical analysis are fitted with log-normal distributions.

\textbf{Raman spectroscopy.}
The graphene inks are drop-cast onto Si/SiO$_2$ wafers (LDB Technologies Ltd.) and dried under vacuum. Raman measurements are collected with a Renishaw inVia confocal Raman microscope using an excitation line of 514\,nm with a 100X objective lens, and an incident power of $\sim 1$\,mW on the sample. 20 spectra are collected for each sample. Peaks are fitted with Lorentzian functions.

\textbf{Scanning electron microscopy (SEM).}
Electrodes are imaged by aim of a field-emission scanning electron microscope FE-SEM (JOEL JSM-7500 FA). The acceleration voltage is set at 5\,kV. Images are collected using the in-lens sensors (secondary electron in-lens image, SEI) and the secondary electron sensor (lower secondary electron image, LEI). No coating is applied.

\textbf{Photoluminescence spectroscopy (PL).}
PL experiments were performed, in a quasi back-scattering geometry, keeping the samples in a closed cycle cryostat and the temperature was changed from 10 to 300\,K. Time integrated photoluminescence (TI-PL) measurements were performed exciting the samples by different mode-locked ps laser sources: a tunable (700--850\,nm and 350--425\,nm with the second harmonic generator) Ti-Sapphire laser operating at 81.3\,MHz repetition rate with 1.2\,ps pulses and a 4\,ps Rhodamine 6G dye laser synchronously pumped by the second harmonic of a mode-locked Nd-YAG laser, operating at 76\,MHz. The PL signal was spectrally dispersed by a 50\,cm monochromator providing a spectral resolution of 1\,meV and detected by a microchannel plate photomultiplier.
Time resolved (TR-PL) measurements were carried out exciting the samples by the ps dye laser operating at 600\,nm and using time-correlated single photon counting technique (TCSPC) with a temporal resolution of about 60\,ps.

\begin{acknowledgments}
FB acknowledges funding from the Italian Ministry for Education, University and Research within the Futuro in Ricerca (FIRB) program (project DeLIGHTeD, Protocollo RBFR12RS1W). We warmly acknowledge Franco Bogani for fruitful discussions. This work was partially supported by ENTE CARIFI grant n.\,2015/11162 and from the European Union's Horizon 2020 research and innovation programme under grant agreement n. 696656 - GrapheneCore1.
\end{acknowledgments}

\bibliography{references}

\begin{thebibliography}{40}%
\makeatletter
\providecommand \@ifxundefined [1]{%
 \@ifx{#1\undefined}
}%
\providecommand \@ifnum [1]{%
 \ifnum #1\expandafter \@firstoftwo
 \else \expandafter \@secondoftwo
 \fi
}%
\providecommand \@ifx [1]{%
 \ifx #1\expandafter \@firstoftwo
 \else \expandafter \@secondoftwo
 \fi
}%
\providecommand \natexlab [1]{#1}%
\providecommand \enquote  [1]{``#1''}%
\providecommand \bibnamefont  [1]{#1}%
\providecommand \bibfnamefont [1]{#1}%
\providecommand \citenamefont [1]{#1}%
\providecommand \href@noop [0]{\@secondoftwo}%
\providecommand \href [0]{\begingroup \@sanitize@url \@href}%
\providecommand \@href[1]{\@@startlink{#1}\@@href}%
\providecommand \@@href[1]{\endgroup#1\@@endlink}%
\providecommand \@sanitize@url [0]{\catcode `\\12\catcode `\$12\catcode
  `\&12\catcode `\#12\catcode `\^12\catcode `\_12\catcode `\%12\relax}%
\providecommand \@@startlink[1]{}%
\providecommand \@@endlink[0]{}%
\providecommand \url  [0]{\begingroup\@sanitize@url \@url }%
\providecommand \@url [1]{\endgroup\@href {#1}{\urlprefix }}%
\providecommand \urlprefix  [0]{URL }%
\providecommand \Eprint [0]{\href }%
\providecommand \doibase [0]{http://dx.doi.org/}%
\providecommand \selectlanguage [0]{\@gobble}%
\providecommand \bibinfo  [0]{\@secondoftwo}%
\providecommand \bibfield  [0]{\@secondoftwo}%
\providecommand \translation [1]{[#1]}%
\providecommand \BibitemOpen [0]{}%
\providecommand \bibitemStop [0]{}%
\providecommand \bibitemNoStop [0]{.\EOS\space}%
\providecommand \EOS [0]{\spacefactor3000\relax}%
\providecommand \BibitemShut  [1]{\csname bibitem#1\endcsname}%
\let\auto@bib@innerbib\@empty
%</preamble>
\bibitem [{\citenamefont {Gr\"atzel}(2014)}]{Graetzel_2014}%
  \BibitemOpen
  \bibfield  {author} {\bibinfo {author} {\bibfnamefont {M.}~\bibnamefont
  {Gr\"atzel}},\ }\href {\doibase 10.1038/nmat4065} {\bibfield  {journal}
  {\bibinfo  {journal} {Nature Materials}\ }\textbf {\bibinfo {volume} {13}},\
  \bibinfo {pages} {838} (\bibinfo {year} {2014})}\BibitemShut {NoStop}%
\bibitem [{\citenamefont {Hodes}(2013)}]{Hodes_2013}%
  \BibitemOpen
  \bibfield  {author} {\bibinfo {author} {\bibfnamefont {G.}~\bibnamefont
  {Hodes}},\ }\href {\doibase 10.1126/science.1245473} {\bibfield  {journal}
  {\bibinfo  {journal} {Science}\ }\textbf {\bibinfo {volume} {342}},\ \bibinfo
  {pages} {317} (\bibinfo {year} {2013})}\BibitemShut {NoStop}%
\bibitem [{\citenamefont {Xiao}\ \emph {et~al.}(2016)\citenamefont {Xiao},
  \citenamefont {Yuan}, \citenamefont {Wang}, \citenamefont {Shao},
  \citenamefont {Bai}, \citenamefont {Deng}, \citenamefont {Dong},
  \citenamefont {Hu}, \citenamefont {Bi},\ and\ \citenamefont
  {Huang}}]{Xiao_2016}%
  \BibitemOpen
  \bibfield  {author} {\bibinfo {author} {\bibfnamefont {Z.}~\bibnamefont
  {Xiao}}, \bibinfo {author} {\bibfnamefont {Y.}~\bibnamefont {Yuan}}, \bibinfo
  {author} {\bibfnamefont {Q.}~\bibnamefont {Wang}}, \bibinfo {author}
  {\bibfnamefont {Y.}~\bibnamefont {Shao}}, \bibinfo {author} {\bibfnamefont
  {Y.}~\bibnamefont {Bai}}, \bibinfo {author} {\bibfnamefont {Y.}~\bibnamefont
  {Deng}}, \bibinfo {author} {\bibfnamefont {Q.}~\bibnamefont {Dong}}, \bibinfo
  {author} {\bibfnamefont {M.}~\bibnamefont {Hu}}, \bibinfo {author}
  {\bibfnamefont {C.}~\bibnamefont {Bi}}, \ and\ \bibinfo {author}
  {\bibfnamefont {J.}~\bibnamefont {Huang}},\ }\href {\doibase
  10.1016/j.mser.2015.12.002} {\bibfield  {journal} {\bibinfo  {journal}
  {Materials Science and Engineering: R: Reports}\ }\textbf {\bibinfo {volume}
  {101}},\ \bibinfo {pages} {1} (\bibinfo {year} {2016})}\BibitemShut {NoStop}%
\bibitem [{\citenamefont {Lee}\ \emph {et~al.}(2012)\citenamefont {Lee},
  \citenamefont {Teuscher}, \citenamefont {Miyasaka}, \citenamefont
  {Murakami},\ and\ \citenamefont {Snaith}}]{Lee_2012}%
  \BibitemOpen
  \bibfield  {author} {\bibinfo {author} {\bibfnamefont {M.~M.}\ \bibnamefont
  {Lee}}, \bibinfo {author} {\bibfnamefont {J.}~\bibnamefont {Teuscher}},
  \bibinfo {author} {\bibfnamefont {T.}~\bibnamefont {Miyasaka}}, \bibinfo
  {author} {\bibfnamefont {T.~N.}\ \bibnamefont {Murakami}}, \ and\ \bibinfo
  {author} {\bibfnamefont {H.~J.}\ \bibnamefont {Snaith}},\ }\href {\doibase
  10.1126/science.1228604} {\bibfield  {journal} {\bibinfo  {journal}
  {Science}\ }\textbf {\bibinfo {volume} {338}},\ \bibinfo {pages} {643}
  (\bibinfo {year} {2012})}\BibitemShut {NoStop}%
\bibitem [{\citenamefont {Heo}\ \emph {et~al.}(2013)\citenamefont {Heo},
  \citenamefont {Im}, \citenamefont {Noh}, \citenamefont {Mandal},
  \citenamefont {Lim}, \citenamefont {Chang}, \citenamefont {Lee},
  \citenamefont {Kim}, \citenamefont {Sarkar}, \citenamefont {Nazeeruddin},
  \citenamefont {Gr\"atzel},\ and\ \citenamefont {Seok}}]{Heo_2013}%
  \BibitemOpen
  \bibfield  {author} {\bibinfo {author} {\bibfnamefont {J.~H.}\ \bibnamefont
  {Heo}}, \bibinfo {author} {\bibfnamefont {S.~H.}\ \bibnamefont {Im}},
  \bibinfo {author} {\bibfnamefont {J.~H.}\ \bibnamefont {Noh}}, \bibinfo
  {author} {\bibfnamefont {T.~N.}\ \bibnamefont {Mandal}}, \bibinfo {author}
  {\bibfnamefont {C.-S.}\ \bibnamefont {Lim}}, \bibinfo {author} {\bibfnamefont
  {J.~A.}\ \bibnamefont {Chang}}, \bibinfo {author} {\bibfnamefont {Y.~H.}\
  \bibnamefont {Lee}}, \bibinfo {author} {\bibfnamefont {H.-j.}\ \bibnamefont
  {Kim}}, \bibinfo {author} {\bibfnamefont {A.}~\bibnamefont {Sarkar}},
  \bibinfo {author} {\bibfnamefont {M.~K.}\ \bibnamefont {Nazeeruddin}},
  \bibinfo {author} {\bibfnamefont {M.}~\bibnamefont {Gr\"atzel}}, \ and\
  \bibinfo {author} {\bibfnamefont {S.~I.}\ \bibnamefont {Seok}},\ }\href
  {\doibase 10.1038/nphoton.2013.80} {\bibfield  {journal} {\bibinfo  {journal}
  {Nature Photonics}\ }\textbf {\bibinfo {volume} {7}},\ \bibinfo {pages} {486}
  (\bibinfo {year} {2013})}\BibitemShut {NoStop}%
\bibitem [{\citenamefont {You}\ \emph {et~al.}(2014)\citenamefont {You},
  \citenamefont {Hong}, \citenamefont {Yang}, \citenamefont {Chen},
  \citenamefont {Cai}, \citenamefont {Song}, \citenamefont {Chen},
  \citenamefont {Lu}, \citenamefont {Liu}, \citenamefont {Zhou},\ and\
  \citenamefont {Yang}}]{You_2014}%
  \BibitemOpen
  \bibfield  {author} {\bibinfo {author} {\bibfnamefont {J.}~\bibnamefont
  {You}}, \bibinfo {author} {\bibfnamefont {Z.}~\bibnamefont {Hong}}, \bibinfo
  {author} {\bibfnamefont {Y.~M.}\ \bibnamefont {Yang}}, \bibinfo {author}
  {\bibfnamefont {Q.}~\bibnamefont {Chen}}, \bibinfo {author} {\bibfnamefont
  {M.}~\bibnamefont {Cai}}, \bibinfo {author} {\bibfnamefont {T.-B.}\
  \bibnamefont {Song}}, \bibinfo {author} {\bibfnamefont {C.-C.}\ \bibnamefont
  {Chen}}, \bibinfo {author} {\bibfnamefont {S.}~\bibnamefont {Lu}}, \bibinfo
  {author} {\bibfnamefont {Y.}~\bibnamefont {Liu}}, \bibinfo {author}
  {\bibfnamefont {H.}~\bibnamefont {Zhou}}, \ and\ \bibinfo {author}
  {\bibfnamefont {Y.}~\bibnamefont {Yang}},\ }\href {\doibase
  10.1021/nn406020d} {\bibfield  {journal} {\bibinfo  {journal} {{ACS} Nano}\
  }\textbf {\bibinfo {volume} {8}},\ \bibinfo {pages} {1674} (\bibinfo {year}
  {2014})}\BibitemShut {NoStop}%
\bibitem [{\citenamefont {Zhou}\ \emph {et~al.}(2014)\citenamefont {Zhou},
  \citenamefont {Chen}, \citenamefont {Li}, \citenamefont {Luo}, \citenamefont
  {Song}, \citenamefont {Duan}, \citenamefont {Hong}, \citenamefont {You},
  \citenamefont {Liu},\ and\ \citenamefont {Yang}}]{Zhou_2014}%
  \BibitemOpen
  \bibfield  {author} {\bibinfo {author} {\bibfnamefont {H.}~\bibnamefont
  {Zhou}}, \bibinfo {author} {\bibfnamefont {Q.}~\bibnamefont {Chen}}, \bibinfo
  {author} {\bibfnamefont {G.}~\bibnamefont {Li}}, \bibinfo {author}
  {\bibfnamefont {S.}~\bibnamefont {Luo}}, \bibinfo {author} {\bibfnamefont
  {T.-b.}\ \bibnamefont {Song}}, \bibinfo {author} {\bibfnamefont {H.-S.}\
  \bibnamefont {Duan}}, \bibinfo {author} {\bibfnamefont {Z.}~\bibnamefont
  {Hong}}, \bibinfo {author} {\bibfnamefont {J.}~\bibnamefont {You}}, \bibinfo
  {author} {\bibfnamefont {Y.}~\bibnamefont {Liu}}, \ and\ \bibinfo {author}
  {\bibfnamefont {Y.}~\bibnamefont {Yang}},\ }\href {\doibase
  10.1126/science.1254050} {\bibfield  {journal} {\bibinfo  {journal}
  {Science}\ }\textbf {\bibinfo {volume} {345}},\ \bibinfo {pages} {542}
  (\bibinfo {year} {2014})}\BibitemShut {NoStop}%
\bibitem [{\citenamefont {Nanni}\ \emph {et~al.}(2014)\citenamefont {Nanni},
  \citenamefont {Lamastra}, \citenamefont {Franceschetti}, \citenamefont
  {Biccari},\ and\ \citenamefont {Cacciotti}}]{Nanni_2014}%
  \BibitemOpen
  \bibfield  {author} {\bibinfo {author} {\bibfnamefont {F.}~\bibnamefont
  {Nanni}}, \bibinfo {author} {\bibfnamefont {F.~R.}\ \bibnamefont {Lamastra}},
  \bibinfo {author} {\bibfnamefont {F.}~\bibnamefont {Franceschetti}}, \bibinfo
  {author} {\bibfnamefont {F.}~\bibnamefont {Biccari}}, \ and\ \bibinfo
  {author} {\bibfnamefont {I.}~\bibnamefont {Cacciotti}},\ }\href {\doibase
  10.1016/j.ceramint.2013.07.087} {\bibfield  {journal} {\bibinfo  {journal}
  {Ceramics International}\ }\textbf {\bibinfo {volume} {40}},\ \bibinfo
  {pages} {1851} (\bibinfo {year} {2014})}\BibitemShut {NoStop}%
\bibitem [{\citenamefont {Kojima}\ \emph {et~al.}(2009)\citenamefont {Kojima},
  \citenamefont {Teshima}, \citenamefont {Shirai},\ and\ \citenamefont
  {Miyasaka}}]{Kojima_2009}%
  \BibitemOpen
  \bibfield  {author} {\bibinfo {author} {\bibfnamefont {A.}~\bibnamefont
  {Kojima}}, \bibinfo {author} {\bibfnamefont {K.}~\bibnamefont {Teshima}},
  \bibinfo {author} {\bibfnamefont {Y.}~\bibnamefont {Shirai}}, \ and\ \bibinfo
  {author} {\bibfnamefont {T.}~\bibnamefont {Miyasaka}},\ }\href {\doibase
  10.1021/ja809598r} {\bibfield  {journal} {\bibinfo  {journal} {J. Am. Chem.
  Soc.}\ }\textbf {\bibinfo {volume} {131}},\ \bibinfo {pages} {6050} (\bibinfo
  {year} {2009})}\BibitemShut {NoStop}%
\bibitem [{NRE(2016)}]{NREL_eff_chart}%
  \BibitemOpen
  \href {http://www.nrel.gov/pv/assets/images/efficiency_chart.jpg} {\enquote
  {\bibinfo {title} {{NREL} best research-cell efficiency chart},}\ } (\bibinfo
  {year} {Nov. 2016})\BibitemShut {NoStop}%
\bibitem [{\citenamefont {Veldhuis}\ \emph {et~al.}(2016)\citenamefont
  {Veldhuis}, \citenamefont {Boix}, \citenamefont {Yantara}, \citenamefont
  {Li}, \citenamefont {Sum}, \citenamefont {Mathews},\ and\ \citenamefont
  {Mhaisalkar}}]{Veldhuis_2016}%
  \BibitemOpen
  \bibfield  {author} {\bibinfo {author} {\bibfnamefont {S.~A.}\ \bibnamefont
  {Veldhuis}}, \bibinfo {author} {\bibfnamefont {P.~P.}\ \bibnamefont {Boix}},
  \bibinfo {author} {\bibfnamefont {N.}~\bibnamefont {Yantara}}, \bibinfo
  {author} {\bibfnamefont {M.}~\bibnamefont {Li}}, \bibinfo {author}
  {\bibfnamefont {T.~C.}\ \bibnamefont {Sum}}, \bibinfo {author} {\bibfnamefont
  {N.}~\bibnamefont {Mathews}}, \ and\ \bibinfo {author} {\bibfnamefont
  {S.~G.}\ \bibnamefont {Mhaisalkar}},\ }\href {\doibase
  10.1002/adma.201600669} {\bibfield  {journal} {\bibinfo  {journal} {Adv.
  Mater.}\ }\textbf {\bibinfo {volume} {28}},\ \bibinfo {pages} {6804}
  (\bibinfo {year} {2016})}\BibitemShut {NoStop}%
\bibitem [{\citenamefont {Sutherland}\ and\ \citenamefont
  {Sargent}(2016)}]{Sutherland_2016}%
  \BibitemOpen
  \bibfield  {author} {\bibinfo {author} {\bibfnamefont {B.~R.}\ \bibnamefont
  {Sutherland}}\ and\ \bibinfo {author} {\bibfnamefont {E.~H.}\ \bibnamefont
  {Sargent}},\ }\href {\doibase 10.1038/nphoton.2016.62} {\bibfield  {journal}
  {\bibinfo  {journal} {Nature Photonics}\ }\textbf {\bibinfo {volume} {10}},\
  \bibinfo {pages} {295} (\bibinfo {year} {2016})}\BibitemShut {NoStop}%
\bibitem [{\citenamefont {Palma}\ \emph
  {et~al.}(2016{\natexlab{a}})\citenamefont {Palma}, \citenamefont
  {Cin{\`{a}}}, \citenamefont {Busby}, \citenamefont {Marsella}, \citenamefont
  {Agresti}, \citenamefont {Pescetelli}, \citenamefont {Pireaux},\ and\
  \citenamefont {Di~Carlo}}]{Palma_2016}%
  \BibitemOpen
  \bibfield  {author} {\bibinfo {author} {\bibfnamefont {A.~L.}\ \bibnamefont
  {Palma}}, \bibinfo {author} {\bibfnamefont {L.}~\bibnamefont {Cin{\`{a}}}},
  \bibinfo {author} {\bibfnamefont {Y.}~\bibnamefont {Busby}}, \bibinfo
  {author} {\bibfnamefont {A.}~\bibnamefont {Marsella}}, \bibinfo {author}
  {\bibfnamefont {A.}~\bibnamefont {Agresti}}, \bibinfo {author} {\bibfnamefont
  {S.}~\bibnamefont {Pescetelli}}, \bibinfo {author} {\bibfnamefont {J.-J.}\
  \bibnamefont {Pireaux}}, \ and\ \bibinfo {author} {\bibfnamefont
  {A.}~\bibnamefont {Di~Carlo}},\ }\href {\doibase 10.1021/acsami.6b07750}
  {\bibfield  {journal} {\bibinfo  {journal} {{ACS} Applied Materials {\&}
  Interfaces}\ }\textbf {\bibinfo {volume} {8}},\ \bibinfo {pages} {26989}
  (\bibinfo {year} {2016}{\natexlab{a}})}\BibitemShut {NoStop}%
\bibitem [{\citenamefont {Palma}\ \emph
  {et~al.}(2016{\natexlab{b}})\citenamefont {Palma}, \citenamefont
  {Cin{\`{a}}}, \citenamefont {Busby}, \citenamefont {Marsella}, \citenamefont
  {Agresti}, \citenamefont {Pescetelli}, \citenamefont {Pireaux},\ and\
  \citenamefont {Di~Carlo}}]{Palma_2016_2}%
  \BibitemOpen
  \bibfield  {author} {\bibinfo {author} {\bibfnamefont {A.~L.}\ \bibnamefont
  {Palma}}, \bibinfo {author} {\bibfnamefont {L.}~\bibnamefont {Cin{\`{a}}}},
  \bibinfo {author} {\bibfnamefont {Y.}~\bibnamefont {Busby}}, \bibinfo
  {author} {\bibfnamefont {A.}~\bibnamefont {Marsella}}, \bibinfo {author}
  {\bibfnamefont {A.}~\bibnamefont {Agresti}}, \bibinfo {author} {\bibfnamefont
  {S.}~\bibnamefont {Pescetelli}}, \bibinfo {author} {\bibfnamefont {J.-J.}\
  \bibnamefont {Pireaux}}, \ and\ \bibinfo {author} {\bibfnamefont
  {A.}~\bibnamefont {Di~Carlo}},\ }\href {\doibase 10.1002/pssc.201600101}
  {\bibfield  {journal} {\bibinfo  {journal} {physica status solidi (c)}\
  }\textbf {\bibinfo {volume} {13}},\ \bibinfo {pages} {958} (\bibinfo {year}
  {2016}{\natexlab{b}})}\BibitemShut {NoStop}%
\bibitem [{\citenamefont {Xu}\ \emph {et~al.}(2016)\citenamefont {Xu},
  \citenamefont {Chen}, \citenamefont {Guo},\ and\ \citenamefont
  {Ma}}]{Xu_2016}%
  \BibitemOpen
  \bibfield  {author} {\bibinfo {author} {\bibfnamefont {T.}~\bibnamefont
  {Xu}}, \bibinfo {author} {\bibfnamefont {L.}~\bibnamefont {Chen}}, \bibinfo
  {author} {\bibfnamefont {Z.}~\bibnamefont {Guo}}, \ and\ \bibinfo {author}
  {\bibfnamefont {T.}~\bibnamefont {Ma}},\ }\href {\doibase 10.1039/c6cp04553g}
  {\bibfield  {journal} {\bibinfo  {journal} {Phys. Chem. Chem. Phys.}\
  }\textbf {\bibinfo {volume} {18}},\ \bibinfo {pages} {27026} (\bibinfo {year}
  {2016})}\BibitemShut {NoStop}%
\bibitem [{\citenamefont {Draguta}\ \emph {et~al.}(2016)\citenamefont
  {Draguta}, \citenamefont {Thakur}, \citenamefont {Morozov}, \citenamefont
  {Wang}, \citenamefont {Manser}, \citenamefont {Kamat},\ and\ \citenamefont
  {Kuno}}]{Draguta_2016}%
  \BibitemOpen
  \bibfield  {author} {\bibinfo {author} {\bibfnamefont {S.}~\bibnamefont
  {Draguta}}, \bibinfo {author} {\bibfnamefont {S.}~\bibnamefont {Thakur}},
  \bibinfo {author} {\bibfnamefont {Y.~V.}\ \bibnamefont {Morozov}}, \bibinfo
  {author} {\bibfnamefont {Y.}~\bibnamefont {Wang}}, \bibinfo {author}
  {\bibfnamefont {J.~S.}\ \bibnamefont {Manser}}, \bibinfo {author}
  {\bibfnamefont {P.~V.}\ \bibnamefont {Kamat}}, \ and\ \bibinfo {author}
  {\bibfnamefont {M.}~\bibnamefont {Kuno}},\ }\href {\doibase
  10.1021/acs.jpclett.5b02888} {\bibfield  {journal} {\bibinfo  {journal} {J.
  Phys. Chem. Lett.}\ }\textbf {\bibinfo {volume} {7}},\ \bibinfo {pages} {715}
  (\bibinfo {year} {2016})}\BibitemShut {NoStop}%
\bibitem [{\citenamefont {Wu}\ \emph {et~al.}(2015)\citenamefont {Wu},
  \citenamefont {Trinh}, \citenamefont {Niesner}, \citenamefont {Zhu},
  \citenamefont {Norman}, \citenamefont {Owen}, \citenamefont {Yaffe},
  \citenamefont {Kudisch},\ and\ \citenamefont {Zhu}}]{Wu_2015}%
  \BibitemOpen
  \bibfield  {author} {\bibinfo {author} {\bibfnamefont {X.}~\bibnamefont
  {Wu}}, \bibinfo {author} {\bibfnamefont {M.~T.}\ \bibnamefont {Trinh}},
  \bibinfo {author} {\bibfnamefont {D.}~\bibnamefont {Niesner}}, \bibinfo
  {author} {\bibfnamefont {H.}~\bibnamefont {Zhu}}, \bibinfo {author}
  {\bibfnamefont {Z.}~\bibnamefont {Norman}}, \bibinfo {author} {\bibfnamefont
  {J.~S.}\ \bibnamefont {Owen}}, \bibinfo {author} {\bibfnamefont
  {O.}~\bibnamefont {Yaffe}}, \bibinfo {author} {\bibfnamefont {B.~J.}\
  \bibnamefont {Kudisch}}, \ and\ \bibinfo {author} {\bibfnamefont {X.-Y.}\
  \bibnamefont {Zhu}},\ }\href {\doibase 10.1021/ja512833n} {\bibfield
  {journal} {\bibinfo  {journal} {J. Am. Chem. Soc.}\ }\textbf {\bibinfo
  {volume} {137}},\ \bibinfo {pages} {2089} (\bibinfo {year}
  {2015})}\BibitemShut {NoStop}%
\bibitem [{\citenamefont {Kim}\ \emph {et~al.}(2014)\citenamefont {Kim},
  \citenamefont {Lee}, \citenamefont {Lee},\ and\ \citenamefont
  {Hong}}]{Kim_2014}%
  \BibitemOpen
  \bibfield  {author} {\bibinfo {author} {\bibfnamefont {J.}~\bibnamefont
  {Kim}}, \bibinfo {author} {\bibfnamefont {S.-H.}\ \bibnamefont {Lee}},
  \bibinfo {author} {\bibfnamefont {J.~H.}\ \bibnamefont {Lee}}, \ and\
  \bibinfo {author} {\bibfnamefont {K.-H.}\ \bibnamefont {Hong}},\ }\href
  {\doibase 10.1021/jz500370k} {\bibfield  {journal} {\bibinfo  {journal} {J.
  Phys. Chem. Lett.}\ }\textbf {\bibinfo {volume} {5}},\ \bibinfo {pages}
  {1312} (\bibinfo {year} {2014})}\BibitemShut {NoStop}%
\bibitem [{\citenamefont {Park}, \citenamefont {Gr\"atzel},\ and\ \citenamefont
  {Miyasaka}(2016)}]{Park_2016_book}%
  \BibitemOpen
  \bibinfo {editor} {\bibfnamefont {N.-G.}\ \bibnamefont {Park}}, \bibinfo
  {editor} {\bibfnamefont {M.}~\bibnamefont {Gr\"atzel}}, \ and\ \bibinfo
  {editor} {\bibfnamefont {T.}~\bibnamefont {Miyasaka}},\ eds.,\ \href
  {\doibase 10.1007/978-3-319-35114-8} {\emph {\bibinfo {title}
  {Organic-Inorganic Halide Perovskite Photovoltaics}}}\ (\bibinfo  {publisher}
  {Springer Nature},\ \bibinfo {year} {2016})\BibitemShut {NoStop}%
\bibitem [{\citenamefont {Son}\ \emph {et~al.}(2016)\citenamefont {Son},
  \citenamefont {Lee}, \citenamefont {Choi}, \citenamefont {Jang},
  \citenamefont {Lee}, \citenamefont {Yoo}, \citenamefont {Shin}, \citenamefont
  {Ahn}, \citenamefont {Choi}, \citenamefont {Kim},\ and\ \citenamefont
  {Park}}]{Son_2016}%
  \BibitemOpen
  \bibfield  {author} {\bibinfo {author} {\bibfnamefont {D.-Y.}\ \bibnamefont
  {Son}}, \bibinfo {author} {\bibfnamefont {J.-W.}\ \bibnamefont {Lee}},
  \bibinfo {author} {\bibfnamefont {Y.~J.}\ \bibnamefont {Choi}}, \bibinfo
  {author} {\bibfnamefont {I.-H.}\ \bibnamefont {Jang}}, \bibinfo {author}
  {\bibfnamefont {S.}~\bibnamefont {Lee}}, \bibinfo {author} {\bibfnamefont
  {P.~J.}\ \bibnamefont {Yoo}}, \bibinfo {author} {\bibfnamefont
  {H.}~\bibnamefont {Shin}}, \bibinfo {author} {\bibfnamefont {N.}~\bibnamefont
  {Ahn}}, \bibinfo {author} {\bibfnamefont {M.}~\bibnamefont {Choi}}, \bibinfo
  {author} {\bibfnamefont {D.}~\bibnamefont {Kim}}, \ and\ \bibinfo {author}
  {\bibfnamefont {N.-G.}\ \bibnamefont {Park}},\ }\href {\doibase
  10.1038/nenergy.2016.81} {\bibfield  {journal} {\bibinfo  {journal} {Nature
  Energy}\ }\textbf {\bibinfo {volume} {1}},\ \bibinfo {pages} {16081}
  (\bibinfo {year} {2016})}\BibitemShut {NoStop}%
\bibitem [{\citenamefont {Giordano}\ \emph {et~al.}(2016)\citenamefont
  {Giordano}, \citenamefont {Abate}, \citenamefont {Baena}, \citenamefont
  {Saliba}, \citenamefont {Matsui}, \citenamefont {Im}, \citenamefont
  {Zakeeruddin}, \citenamefont {Nazeeruddin}, \citenamefont {Hagfeldt},\ and\
  \citenamefont {Gr\"atzel}}]{Giordano_2016}%
  \BibitemOpen
  \bibfield  {author} {\bibinfo {author} {\bibfnamefont {F.}~\bibnamefont
  {Giordano}}, \bibinfo {author} {\bibfnamefont {A.}~\bibnamefont {Abate}},
  \bibinfo {author} {\bibfnamefont {J.~P.~C.}\ \bibnamefont {Baena}}, \bibinfo
  {author} {\bibfnamefont {M.}~\bibnamefont {Saliba}}, \bibinfo {author}
  {\bibfnamefont {T.}~\bibnamefont {Matsui}}, \bibinfo {author} {\bibfnamefont
  {S.~H.}\ \bibnamefont {Im}}, \bibinfo {author} {\bibfnamefont {S.~M.}\
  \bibnamefont {Zakeeruddin}}, \bibinfo {author} {\bibfnamefont {M.~K.}\
  \bibnamefont {Nazeeruddin}}, \bibinfo {author} {\bibfnamefont
  {A.}~\bibnamefont {Hagfeldt}}, \ and\ \bibinfo {author} {\bibfnamefont
  {M.}~\bibnamefont {Gr\"atzel}},\ }\href {\doibase 10.1038/ncomms10379}
  {\bibfield  {journal} {\bibinfo  {journal} {Nature Communications}\ }\textbf
  {\bibinfo {volume} {7}},\ \bibinfo {pages} {10379} (\bibinfo {year}
  {2016})}\BibitemShut {NoStop}%
\bibitem [{\citenamefont {deQuilettes}\ \emph {et~al.}(2016)\citenamefont
  {deQuilettes}, \citenamefont {Zhang}, \citenamefont {Burlakov}, \citenamefont
  {Graham}, \citenamefont {Leijtens}, \citenamefont {Osherov}, \citenamefont
  {Bulovi{\'{c}}}, \citenamefont {Snaith}, \citenamefont {Ginger},\ and\
  \citenamefont {Stranks}}]{deQuilettes_2016}%
  \BibitemOpen
  \bibfield  {author} {\bibinfo {author} {\bibfnamefont {D.~W.}\ \bibnamefont
  {deQuilettes}}, \bibinfo {author} {\bibfnamefont {W.}~\bibnamefont {Zhang}},
  \bibinfo {author} {\bibfnamefont {V.~M.}\ \bibnamefont {Burlakov}}, \bibinfo
  {author} {\bibfnamefont {D.~J.}\ \bibnamefont {Graham}}, \bibinfo {author}
  {\bibfnamefont {T.}~\bibnamefont {Leijtens}}, \bibinfo {author}
  {\bibfnamefont {A.}~\bibnamefont {Osherov}}, \bibinfo {author} {\bibfnamefont
  {V.}~\bibnamefont {Bulovi{\'{c}}}}, \bibinfo {author} {\bibfnamefont {H.~J.}\
  \bibnamefont {Snaith}}, \bibinfo {author} {\bibfnamefont {D.~S.}\
  \bibnamefont {Ginger}}, \ and\ \bibinfo {author} {\bibfnamefont {S.~D.}\
  \bibnamefont {Stranks}},\ }\href {\doibase 10.1038/ncomms11683} {\bibfield
  {journal} {\bibinfo  {journal} {Nature Communications}\ }\textbf {\bibinfo
  {volume} {7}},\ \bibinfo {pages} {11683} (\bibinfo {year}
  {2016})}\BibitemShut {NoStop}%
\bibitem [{\citenamefont {Shao}, \citenamefont {Yuan},\ and\ \citenamefont
  {Huang}(2016)}]{Shao_2016}%
  \BibitemOpen
  \bibfield  {author} {\bibinfo {author} {\bibfnamefont {Y.}~\bibnamefont
  {Shao}}, \bibinfo {author} {\bibfnamefont {Y.}~\bibnamefont {Yuan}}, \ and\
  \bibinfo {author} {\bibfnamefont {J.}~\bibnamefont {Huang}},\ }\href
  {\doibase 10.1038/nenergy.2015.1} {\bibfield  {journal} {\bibinfo  {journal}
  {Nature Energy}\ }\textbf {\bibinfo {volume} {1}},\ \bibinfo {pages} {15001}
  (\bibinfo {year} {2016})}\BibitemShut {NoStop}%
\bibitem [{\citenamefont {Leblebici}\ \emph {et~al.}(2016)\citenamefont
  {Leblebici}, \citenamefont {Leppert}, \citenamefont {Li}, \citenamefont
  {Reyes-Lillo}, \citenamefont {Wickenburg}, \citenamefont {Wong},
  \citenamefont {Lee}, \citenamefont {Melli}, \citenamefont {Ziegler},
  \citenamefont {Angell}, \citenamefont {Ogletree}, \citenamefont {Ashby},
  \citenamefont {Toma}, \citenamefont {Neaton}, \citenamefont {Sharp},\ and\
  \citenamefont {Weber-Bargioni}}]{Leblebici_2016}%
  \BibitemOpen
  \bibfield  {author} {\bibinfo {author} {\bibfnamefont {S.~Y.}\ \bibnamefont
  {Leblebici}}, \bibinfo {author} {\bibfnamefont {L.}~\bibnamefont {Leppert}},
  \bibinfo {author} {\bibfnamefont {Y.}~\bibnamefont {Li}}, \bibinfo {author}
  {\bibfnamefont {S.~E.}\ \bibnamefont {Reyes-Lillo}}, \bibinfo {author}
  {\bibfnamefont {S.}~\bibnamefont {Wickenburg}}, \bibinfo {author}
  {\bibfnamefont {E.}~\bibnamefont {Wong}}, \bibinfo {author} {\bibfnamefont
  {J.}~\bibnamefont {Lee}}, \bibinfo {author} {\bibfnamefont {M.}~\bibnamefont
  {Melli}}, \bibinfo {author} {\bibfnamefont {D.}~\bibnamefont {Ziegler}},
  \bibinfo {author} {\bibfnamefont {D.~K.}\ \bibnamefont {Angell}}, \bibinfo
  {author} {\bibfnamefont {D.~F.}\ \bibnamefont {Ogletree}}, \bibinfo {author}
  {\bibfnamefont {P.~D.}\ \bibnamefont {Ashby}}, \bibinfo {author}
  {\bibfnamefont {F.~M.}\ \bibnamefont {Toma}}, \bibinfo {author}
  {\bibfnamefont {J.~B.}\ \bibnamefont {Neaton}}, \bibinfo {author}
  {\bibfnamefont {I.~D.}\ \bibnamefont {Sharp}}, \ and\ \bibinfo {author}
  {\bibfnamefont {A.}~\bibnamefont {Weber-Bargioni}},\ }\href {\doibase
  10.1038/nenergy.2016.93} {\bibfield  {journal} {\bibinfo  {journal} {Nature
  Energy}\ }\textbf {\bibinfo {volume} {1}},\ \bibinfo {pages} {16093 EP}
  (\bibinfo {year} {2016})}\BibitemShut {NoStop}%
\bibitem [{\citenamefont {Acik}\ and\ \citenamefont
  {Darling}(2016)}]{Acik_2016}%
  \BibitemOpen
  \bibfield  {author} {\bibinfo {author} {\bibfnamefont {M.}~\bibnamefont
  {Acik}}\ and\ \bibinfo {author} {\bibfnamefont {S.~B.}\ \bibnamefont
  {Darling}},\ }\href {\doibase 10.1039/c5ta09911k} {\bibfield  {journal}
  {\bibinfo  {journal} {J. Mater. Chem. A}\ }\textbf {\bibinfo {volume} {4}},\
  \bibinfo {pages} {6185} (\bibinfo {year} {2016})}\BibitemShut {NoStop}%
\bibitem [{\citenamefont {Agresti}\ \emph
  {et~al.}(2016{\natexlab{a}})\citenamefont {Agresti}, \citenamefont
  {Pescetelli}, \citenamefont {Taheri}, \citenamefont {Del Rio~Castillo},
  \citenamefont {Cin{\`{a}}}, \citenamefont {Bonaccorso},\ and\ \citenamefont
  {Di~Carlo}}]{Agresti_2016}%
  \BibitemOpen
  \bibfield  {author} {\bibinfo {author} {\bibfnamefont {A.}~\bibnamefont
  {Agresti}}, \bibinfo {author} {\bibfnamefont {S.}~\bibnamefont {Pescetelli}},
  \bibinfo {author} {\bibfnamefont {B.}~\bibnamefont {Taheri}}, \bibinfo
  {author} {\bibfnamefont {A.~E.}\ \bibnamefont {Del Rio~Castillo}}, \bibinfo
  {author} {\bibfnamefont {L.}~\bibnamefont {Cin{\`{a}}}}, \bibinfo {author}
  {\bibfnamefont {F.}~\bibnamefont {Bonaccorso}}, \ and\ \bibinfo {author}
  {\bibfnamefont {A.}~\bibnamefont {Di~Carlo}},\ }\href {\doibase
  10.1002/cssc.201600942} {\bibfield  {journal} {\bibinfo  {journal}
  {{ChemSusChem}}\ }\textbf {\bibinfo {volume} {9}},\ \bibinfo {pages} {2609}
  (\bibinfo {year} {2016}{\natexlab{a}})}\BibitemShut {NoStop}%
\bibitem [{\citenamefont {Capasso}\ \emph {et~al.}(2016)\citenamefont
  {Capasso}, \citenamefont {Matteocci}, \citenamefont {Najafi}, \citenamefont
  {Prato}, \citenamefont {Buha}, \citenamefont {Cin{\`{a}}}, \citenamefont
  {Pellegrini}, \citenamefont {Carlo},\ and\ \citenamefont
  {Bonaccorso}}]{Capasso_2016}%
  \BibitemOpen
  \bibfield  {author} {\bibinfo {author} {\bibfnamefont {A.}~\bibnamefont
  {Capasso}}, \bibinfo {author} {\bibfnamefont {F.}~\bibnamefont {Matteocci}},
  \bibinfo {author} {\bibfnamefont {L.}~\bibnamefont {Najafi}}, \bibinfo
  {author} {\bibfnamefont {M.}~\bibnamefont {Prato}}, \bibinfo {author}
  {\bibfnamefont {J.}~\bibnamefont {Buha}}, \bibinfo {author} {\bibfnamefont
  {L.}~\bibnamefont {Cin{\`{a}}}}, \bibinfo {author} {\bibfnamefont
  {V.}~\bibnamefont {Pellegrini}}, \bibinfo {author} {\bibfnamefont {A.~D.}\
  \bibnamefont {Carlo}}, \ and\ \bibinfo {author} {\bibfnamefont
  {F.}~\bibnamefont {Bonaccorso}},\ }\href {\doibase 10.1002/aenm.201600920}
  {\bibfield  {journal} {\bibinfo  {journal} {Advanced Energy Materials}\
  }\textbf {\bibinfo {volume} {6}},\ \bibinfo {pages} {1600920} (\bibinfo
  {year} {2016})}\BibitemShut {NoStop}%
\bibitem [{\citenamefont {Ahn}\ \emph {et~al.}(2016)\citenamefont {Ahn},
  \citenamefont {Kwak}, \citenamefont {Jang}, \citenamefont {Yoon},
  \citenamefont {Yang}, \citenamefont {Lee}, \citenamefont {Pikhitsa},
  \citenamefont {Byun},\ and\ \citenamefont {Choi}}]{Ahn_2016}%
  \BibitemOpen
  \bibfield  {author} {\bibinfo {author} {\bibfnamefont {N.}~\bibnamefont
  {Ahn}}, \bibinfo {author} {\bibfnamefont {K.}~\bibnamefont {Kwak}}, \bibinfo
  {author} {\bibfnamefont {M.~S.}\ \bibnamefont {Jang}}, \bibinfo {author}
  {\bibfnamefont {H.}~\bibnamefont {Yoon}}, \bibinfo {author} {\bibfnamefont
  {B.}~\bibnamefont {Yang}}, \bibinfo {author} {\bibfnamefont {J.}~\bibnamefont
  {Lee}}, \bibinfo {author} {\bibfnamefont {P.~V.}\ \bibnamefont {Pikhitsa}},
  \bibinfo {author} {\bibfnamefont {J.}~\bibnamefont {Byun}}, \ and\ \bibinfo
  {author} {\bibfnamefont {M.}~\bibnamefont {Choi}},\ }\href {\doibase
  10.1038/ncomms13422} {\bibfield  {journal} {\bibinfo  {journal} {Nature
  Communication}\ }\textbf {\bibinfo {volume} {7}},\ \bibinfo {pages} {13422}
  (\bibinfo {year} {2016})}\BibitemShut {NoStop}%
\bibitem [{\citenamefont {Agresti}\ \emph {et~al.}(2017)\citenamefont
  {Agresti}, \citenamefont {Pescetelli}, \citenamefont {Palma}, \citenamefont
  {Castillo}, \citenamefont {Konios}, \citenamefont {Kakavelakis},
  \citenamefont {Razza}, \citenamefont {Cin{\`{a}}}, \citenamefont {Kymakis},
  \citenamefont {Bonaccorso},\ and\ \citenamefont {Carlo}}]{Agresti_2017}%
  \BibitemOpen
  \bibfield  {author} {\bibinfo {author} {\bibfnamefont {A.}~\bibnamefont
  {Agresti}}, \bibinfo {author} {\bibfnamefont {S.}~\bibnamefont {Pescetelli}},
  \bibinfo {author} {\bibfnamefont {A.~L.}\ \bibnamefont {Palma}}, \bibinfo
  {author} {\bibfnamefont {A.~E. D.~R.}\ \bibnamefont {Castillo}}, \bibinfo
  {author} {\bibfnamefont {D.}~\bibnamefont {Konios}}, \bibinfo {author}
  {\bibfnamefont {G.}~\bibnamefont {Kakavelakis}}, \bibinfo {author}
  {\bibfnamefont {S.}~\bibnamefont {Razza}}, \bibinfo {author} {\bibfnamefont
  {L.}~\bibnamefont {Cin{\`{a}}}}, \bibinfo {author} {\bibfnamefont
  {E.}~\bibnamefont {Kymakis}}, \bibinfo {author} {\bibfnamefont
  {F.}~\bibnamefont {Bonaccorso}}, \ and\ \bibinfo {author} {\bibfnamefont
  {A.~D.}\ \bibnamefont {Carlo}},\ }\href {\doibase
  10.1021/acsenergylett.6b00672} {\bibfield  {journal} {\bibinfo  {journal}
  {{ACS} Energy Letters}\ }\textbf {\bibinfo {volume} {2}},\ \bibinfo {pages}
  {279} (\bibinfo {year} {2017})}\BibitemShut {NoStop}%
\bibitem [{\citenamefont {L\"oper}\ \emph {et~al.}(2015)\citenamefont
  {L\"oper}, \citenamefont {Stuckelberger}, \citenamefont {Niesen},
  \citenamefont {Werner}, \citenamefont {Filipi{\v{c}}}, \citenamefont {Moon},
  \citenamefont {Yum}, \citenamefont {Topi{\v{c}}}, \citenamefont {Wolf},\ and\
  \citenamefont {Ballif}}]{Loper_2015}%
  \BibitemOpen
  \bibfield  {author} {\bibinfo {author} {\bibfnamefont {P.}~\bibnamefont
  {L\"oper}}, \bibinfo {author} {\bibfnamefont {M.}~\bibnamefont
  {Stuckelberger}}, \bibinfo {author} {\bibfnamefont {B.}~\bibnamefont
  {Niesen}}, \bibinfo {author} {\bibfnamefont {J.}~\bibnamefont {Werner}},
  \bibinfo {author} {\bibfnamefont {M.}~\bibnamefont {Filipi{\v{c}}}}, \bibinfo
  {author} {\bibfnamefont {S.-J.}\ \bibnamefont {Moon}}, \bibinfo {author}
  {\bibfnamefont {J.-H.}\ \bibnamefont {Yum}}, \bibinfo {author} {\bibfnamefont
  {M.}~\bibnamefont {Topi{\v{c}}}}, \bibinfo {author} {\bibfnamefont {S.~D.}\
  \bibnamefont {Wolf}}, \ and\ \bibinfo {author} {\bibfnamefont
  {C.}~\bibnamefont {Ballif}},\ }\href {\doibase 10.1021/jz502471h} {\bibfield
  {journal} {\bibinfo  {journal} {The Journal of Physical Chemistry Letters}\
  }\textbf {\bibinfo {volume} {6}},\ \bibinfo {pages} {66} (\bibinfo {year}
  {2015})}\BibitemShut {NoStop}%
\bibitem [{\citenamefont {Milot}\ \emph {et~al.}(2015)\citenamefont {Milot},
  \citenamefont {Eperon}, \citenamefont {Snaith}, \citenamefont {Johnston},\
  and\ \citenamefont {Herz}}]{Milot_2015}%
  \BibitemOpen
  \bibfield  {author} {\bibinfo {author} {\bibfnamefont {R.~L.}\ \bibnamefont
  {Milot}}, \bibinfo {author} {\bibfnamefont {G.~E.}\ \bibnamefont {Eperon}},
  \bibinfo {author} {\bibfnamefont {H.~J.}\ \bibnamefont {Snaith}}, \bibinfo
  {author} {\bibfnamefont {M.~B.}\ \bibnamefont {Johnston}}, \ and\ \bibinfo
  {author} {\bibfnamefont {L.~M.}\ \bibnamefont {Herz}},\ }\href {\doibase
  10.1002/adfm.201502340} {\bibfield  {journal} {\bibinfo  {journal} {Advanced
  Functional Materials}\ }\textbf {\bibinfo {volume} {25}},\ \bibinfo {pages}
  {6218} (\bibinfo {year} {2015})}\BibitemShut {NoStop}%
\bibitem [{\citenamefont {Dar}\ \emph {et~al.}(2016)\citenamefont {Dar},
  \citenamefont {Jacopin}, \citenamefont {Meloni}, \citenamefont {Mattoni},
  \citenamefont {Arora}, \citenamefont {Boziki}, \citenamefont {Zakeeruddin},
  \citenamefont {Rothlisberger},\ and\ \citenamefont {tzel}}]{Dar_2016}%
  \BibitemOpen
  \bibfield  {author} {\bibinfo {author} {\bibfnamefont {M.~I.}\ \bibnamefont
  {Dar}}, \bibinfo {author} {\bibfnamefont {G.}~\bibnamefont {Jacopin}},
  \bibinfo {author} {\bibfnamefont {S.}~\bibnamefont {Meloni}}, \bibinfo
  {author} {\bibfnamefont {A.}~\bibnamefont {Mattoni}}, \bibinfo {author}
  {\bibfnamefont {N.}~\bibnamefont {Arora}}, \bibinfo {author} {\bibfnamefont
  {A.}~\bibnamefont {Boziki}}, \bibinfo {author} {\bibfnamefont {S.~M.}\
  \bibnamefont {Zakeeruddin}}, \bibinfo {author} {\bibfnamefont
  {U.}~\bibnamefont {Rothlisberger}}, \ and\ \bibinfo {author} {\bibfnamefont
  {M.~G.}\ \bibnamefont {tzel}},\ }\href {\doibase 10.1126/sciadv.1601156}
  {\bibfield  {journal} {\bibinfo  {journal} {Science Advances}\ }\textbf
  {\bibinfo {volume} {2}},\ \bibinfo {pages} {e1601156} (\bibinfo {year}
  {2016})}\BibitemShut {NoStop}%
\bibitem [{\citenamefont {Bi}\ \emph {et~al.}(2016)\citenamefont {Bi},
  \citenamefont {Hutter}, \citenamefont {Fang}, \citenamefont {Dong},
  \citenamefont {Huang},\ and\ \citenamefont {Savenije}}]{Bi_2016}%
  \BibitemOpen
  \bibfield  {author} {\bibinfo {author} {\bibfnamefont {Y.}~\bibnamefont
  {Bi}}, \bibinfo {author} {\bibfnamefont {E.~M.}\ \bibnamefont {Hutter}},
  \bibinfo {author} {\bibfnamefont {Y.}~\bibnamefont {Fang}}, \bibinfo {author}
  {\bibfnamefont {Q.}~\bibnamefont {Dong}}, \bibinfo {author} {\bibfnamefont
  {J.}~\bibnamefont {Huang}}, \ and\ \bibinfo {author} {\bibfnamefont {T.~J.}\
  \bibnamefont {Savenije}},\ }\href {\doibase 10.1021/acs.jpclett.6b00269}
  {\bibfield  {journal} {\bibinfo  {journal} {The Journal of Physical Chemistry
  Letters}\ }\textbf {\bibinfo {volume} {7}},\ \bibinfo {pages} {923} (\bibinfo
  {year} {2016})}\BibitemShut {NoStop}%
\bibitem [{\citenamefont {Warren}\ \emph {et~al.}(2013)\citenamefont {Warren},
  \citenamefont {Margineanu}, \citenamefont {Alibhai}, \citenamefont {Kelly},
  \citenamefont {Talbot}, \citenamefont {Alexandrov}, \citenamefont {Munro},
  \citenamefont {Katan}, \citenamefont {Dunsby},\ and\ \citenamefont
  {French}}]{Warren_2013}%
  \BibitemOpen
  \bibfield  {author} {\bibinfo {author} {\bibfnamefont {S.~C.}\ \bibnamefont
  {Warren}}, \bibinfo {author} {\bibfnamefont {A.}~\bibnamefont {Margineanu}},
  \bibinfo {author} {\bibfnamefont {D.}~\bibnamefont {Alibhai}}, \bibinfo
  {author} {\bibfnamefont {D.~J.}\ \bibnamefont {Kelly}}, \bibinfo {author}
  {\bibfnamefont {C.}~\bibnamefont {Talbot}}, \bibinfo {author} {\bibfnamefont
  {Y.}~\bibnamefont {Alexandrov}}, \bibinfo {author} {\bibfnamefont
  {I.}~\bibnamefont {Munro}}, \bibinfo {author} {\bibfnamefont
  {M.}~\bibnamefont {Katan}}, \bibinfo {author} {\bibfnamefont
  {C.}~\bibnamefont {Dunsby}}, \ and\ \bibinfo {author} {\bibfnamefont
  {P.~M.~W.}\ \bibnamefont {French}},\ }\href {\doibase
  10.1371/journal.pone.0070687} {\bibfield  {journal} {\bibinfo  {journal}
  {{PLoS} {ONE}}\ }\textbf {\bibinfo {volume} {8}},\ \bibinfo {pages} {e70687}
  (\bibinfo {year} {2013})}\BibitemShut {NoStop}%
\bibitem [{\citenamefont {Pydzi{\'{n}}ska}\ \emph {et~al.}(2016)\citenamefont
  {Pydzi{\'{n}}ska}, \citenamefont {Karolczak}, \citenamefont {Kosta},
  \citenamefont {Tena-Zaera}, \citenamefont {Todinova}, \citenamefont
  {Id{\'{\i}}goras}, \citenamefont {Anta},\ and\ \citenamefont
  {Zi{\'{o}}{\l}ek}}]{Pydzinska_2016}%
  \BibitemOpen
  \bibfield  {author} {\bibinfo {author} {\bibfnamefont {K.}~\bibnamefont
  {Pydzi{\'{n}}ska}}, \bibinfo {author} {\bibfnamefont {J.}~\bibnamefont
  {Karolczak}}, \bibinfo {author} {\bibfnamefont {I.}~\bibnamefont {Kosta}},
  \bibinfo {author} {\bibfnamefont {R.}~\bibnamefont {Tena-Zaera}}, \bibinfo
  {author} {\bibfnamefont {A.}~\bibnamefont {Todinova}}, \bibinfo {author}
  {\bibfnamefont {J.}~\bibnamefont {Id{\'{\i}}goras}}, \bibinfo {author}
  {\bibfnamefont {J.~A.}\ \bibnamefont {Anta}}, \ and\ \bibinfo {author}
  {\bibfnamefont {M.}~\bibnamefont {Zi{\'{o}}{\l}ek}},\ }\href {\doibase
  10.1002/cssc.201600210} {\bibfield  {journal} {\bibinfo  {journal}
  {{ChemSusChem}}\ }\textbf {\bibinfo {volume} {9}},\ \bibinfo {pages} {1647}
  (\bibinfo {year} {2016})}\BibitemShut {NoStop}%
\bibitem [{\citenamefont {Kong}\ \emph {et~al.}(2015)\citenamefont {Kong},
  \citenamefont {Ye}, \citenamefont {Qi}, \citenamefont {Zhang}, \citenamefont
  {Wang}, \citenamefont {Rahimi-Iman},\ and\ \citenamefont {Wu}}]{Kong_2015}%
  \BibitemOpen
  \bibfield  {author} {\bibinfo {author} {\bibfnamefont {W.}~\bibnamefont
  {Kong}}, \bibinfo {author} {\bibfnamefont {Z.}~\bibnamefont {Ye}}, \bibinfo
  {author} {\bibfnamefont {Z.}~\bibnamefont {Qi}}, \bibinfo {author}
  {\bibfnamefont {B.}~\bibnamefont {Zhang}}, \bibinfo {author} {\bibfnamefont
  {M.}~\bibnamefont {Wang}}, \bibinfo {author} {\bibfnamefont {A.}~\bibnamefont
  {Rahimi-Iman}}, \ and\ \bibinfo {author} {\bibfnamefont {H.}~\bibnamefont
  {Wu}},\ }\href {\doibase 10.1039/c5cp02605a} {\bibfield  {journal} {\bibinfo
  {journal} {Phys. Chem. Chem. Phys.}\ }\textbf {\bibinfo {volume} {17}},\
  \bibinfo {pages} {16405} (\bibinfo {year} {2015})}\BibitemShut {NoStop}%
\bibitem [{\citenamefont {Fang}\ \emph {et~al.}(2015)\citenamefont {Fang},
  \citenamefont {Raissa}, \citenamefont {Abdu-Aguye}, \citenamefont
  {Adjokatse}, \citenamefont {Blake}, \citenamefont {Even},\ and\ \citenamefont
  {Loi}}]{Fang_2015}%
  \BibitemOpen
  \bibfield  {author} {\bibinfo {author} {\bibfnamefont {H.-H.}\ \bibnamefont
  {Fang}}, \bibinfo {author} {\bibfnamefont {R.}~\bibnamefont {Raissa}},
  \bibinfo {author} {\bibfnamefont {M.}~\bibnamefont {Abdu-Aguye}}, \bibinfo
  {author} {\bibfnamefont {S.}~\bibnamefont {Adjokatse}}, \bibinfo {author}
  {\bibfnamefont {G.~R.}\ \bibnamefont {Blake}}, \bibinfo {author}
  {\bibfnamefont {J.}~\bibnamefont {Even}}, \ and\ \bibinfo {author}
  {\bibfnamefont {M.~A.}\ \bibnamefont {Loi}},\ }\href {\doibase
  10.1002/adfm.201404421} {\bibfield  {journal} {\bibinfo  {journal} {Advanced
  Functional Materials}\ }\textbf {\bibinfo {volume} {25}},\ \bibinfo {pages}
  {2378} (\bibinfo {year} {2015})}\BibitemShut {NoStop}%
\bibitem [{\citenamefont {Wright}\ \emph {et~al.}(2016)\citenamefont {Wright},
  \citenamefont {Verdi}, \citenamefont {Milot}, \citenamefont {Eperon},
  \citenamefont {P{\'{e}}rez-Osorio}, \citenamefont {Snaith}, \citenamefont
  {Giustino}, \citenamefont {Johnston},\ and\ \citenamefont
  {Herz}}]{Wright_2016}%
  \BibitemOpen
  \bibfield  {author} {\bibinfo {author} {\bibfnamefont {A.~D.}\ \bibnamefont
  {Wright}}, \bibinfo {author} {\bibfnamefont {C.}~\bibnamefont {Verdi}},
  \bibinfo {author} {\bibfnamefont {R.~L.}\ \bibnamefont {Milot}}, \bibinfo
  {author} {\bibfnamefont {G.~E.}\ \bibnamefont {Eperon}}, \bibinfo {author}
  {\bibfnamefont {M.~A.}\ \bibnamefont {P{\'{e}}rez-Osorio}}, \bibinfo {author}
  {\bibfnamefont {H.~J.}\ \bibnamefont {Snaith}}, \bibinfo {author}
  {\bibfnamefont {F.}~\bibnamefont {Giustino}}, \bibinfo {author}
  {\bibfnamefont {M.~B.}\ \bibnamefont {Johnston}}, \ and\ \bibinfo {author}
  {\bibfnamefont {L.~M.}\ \bibnamefont {Herz}},\ }\href {\doibase
  10.1038/ncomms11755} {\bibfield  {journal} {\bibinfo  {journal} {Nature
  Communications}\ }\textbf {\bibinfo {volume} {7}},\ \bibinfo {pages} {11755}
  (\bibinfo {year} {2016})}\BibitemShut {NoStop}%
\bibitem [{\citenamefont {Agresti}\ \emph
  {et~al.}(2016{\natexlab{b}})\citenamefont {Agresti}, \citenamefont
  {Pescetelli}, \citenamefont {Cin{\`{a}}}, \citenamefont {Konios},
  \citenamefont {Kakavelakis}, \citenamefont {Kymakis},\ and\ \citenamefont
  {Di~Carlo}}]{Agresti_2016b}%
  \BibitemOpen
  \bibfield  {author} {\bibinfo {author} {\bibfnamefont {A.}~\bibnamefont
  {Agresti}}, \bibinfo {author} {\bibfnamefont {S.}~\bibnamefont {Pescetelli}},
  \bibinfo {author} {\bibfnamefont {L.}~\bibnamefont {Cin{\`{a}}}}, \bibinfo
  {author} {\bibfnamefont {D.}~\bibnamefont {Konios}}, \bibinfo {author}
  {\bibfnamefont {G.}~\bibnamefont {Kakavelakis}}, \bibinfo {author}
  {\bibfnamefont {E.}~\bibnamefont {Kymakis}}, \ and\ \bibinfo {author}
  {\bibfnamefont {A.}~\bibnamefont {Di~Carlo}},\ }\href {\doibase
  10.1002/adfm.201504949} {\bibfield  {journal} {\bibinfo  {journal} {Advanced
  Functional Materials}\ }\textbf {\bibinfo {volume} {26}},\ \bibinfo {pages}
  {2686} (\bibinfo {year} {2016}{\natexlab{b}})}\BibitemShut {NoStop}%
\bibitem [{\citenamefont {Lotya}\ \emph {et~al.}(2009)\citenamefont {Lotya},
  \citenamefont {Hernandez}, \citenamefont {King}, \citenamefont {Smith},
  \citenamefont {Nicolosi}, \citenamefont {Karlsson}, \citenamefont {Blighe},
  \citenamefont {Zhiming}, \citenamefont {McGovern}, \citenamefont {Duesberg},\
  and\ \citenamefont {Coleman}}]{Lotya_2009}%
  \BibitemOpen
  \bibfield  {author} {\bibinfo {author} {\bibfnamefont {M.}~\bibnamefont
  {Lotya}}, \bibinfo {author} {\bibfnamefont {Y.}~\bibnamefont {Hernandez}},
  \bibinfo {author} {\bibfnamefont {P.~J.}\ \bibnamefont {King}}, \bibinfo
  {author} {\bibfnamefont {R.~J.}\ \bibnamefont {Smith}}, \bibinfo {author}
  {\bibfnamefont {V.}~\bibnamefont {Nicolosi}}, \bibinfo {author}
  {\bibfnamefont {L.~S.}\ \bibnamefont {Karlsson}}, \bibinfo {author}
  {\bibfnamefont {F.~M.}\ \bibnamefont {Blighe}}, \bibinfo {author}
  {\bibfnamefont {S.~D.}\ \bibnamefont {Zhiming}}, \bibinfo {author}
  {\bibfnamefont {L.~T.}\ \bibnamefont {McGovern}}, \bibinfo {author}
  {\bibfnamefont {G.~S.}\ \bibnamefont {Duesberg}}, \ and\ \bibinfo {author}
  {\bibfnamefont {J.~N.}\ \bibnamefont {Coleman}},\ }\href@noop {} {\bibfield
  {journal} {\bibinfo  {journal} {J. Am. Chem Soc.}\ }\textbf {\bibinfo
  {volume} {131}},\ \bibinfo {pages} {3611} (\bibinfo {year}
  {2009})}\BibitemShut {NoStop}%
\end{thebibliography}%

\newpage

\printfigures
\newpage
\printtables
\end{document}